%% file: singlev2.tex
%\input psfig
%%%%%%%%%%%%%%%%%%%%%%%%%%%%%%%%%%%%%%% 
\input myharvmac
%%%%%%%%%%%%%%%%%%%%%%%%%%%%%%%%%%%%%%%%

\noblackbox
%%\draftmode  %%%%%%%%%%%%%%%% Lineskip
%%%%%%%%%%%%%%%%%%%%%%%%%%%%%%%%%%%%%%%
\baselineskip=14.5pt
%%%%%%%%%%%% Local definitons %%%%%%%%%%%%%%%%%%%%%%%%%%%%%%%%%%%%

\def\comment#1{{}}
\def\ss#1{{\scriptstyle{#1}}}

\def\z{{\zeta}}
\def\ap{\alpha'}

\def\cf{{\it cf.\ }}
\def\ie{{\it i.e.\ }}

\def\eqq{{\it Eq.\ }}
\def\eqqs{{\it Eqs.\ }}

\def\Hc{{\cal H}}
\def\Uc{{\cal U}}
\def\Qc{{\cal Q}}

%%%%%%%%%%%%%%%%%%%%%%%%%%%%%%%%%%%%%%%%%%%%%%%%%%%%%%%%%%%%%%%%%
%%%%% Referencing  %%%%%%%%%%%%%%%%%%%%%%%%%%%%%%%%%%%%%%%%%%%%%%
%%%%%%%%%%%%%%%%%%%%%%%%%%%%%%%%%%%%%%%%%%%%%%%%%%%%%%%%%%%%%%%%%
\newif\ifnref

\def\doubref#1#2{\refs{{#1},{#2} }}

\nreffalse
%%%%%%%%%%%%%%%%%%%%%%%%%%%%%%%%%%%%%%%%%%%%%%%%%%%%%%%%%%%%%%%%%%
%%%%%%%%%%%%%%%%%   Stuff for Figures  %%%%%%%%%%%%%%%%%%%%%%%%%%%
%%%%%%%%%%%%%%%%%%%%%%%%%%%%%%%%%%%%%%%%%%%%%%%%%%%%%%%%%%%%%%%%%%

\input epsf

\def\figin{\epsfcheck\figin}\def\figins{\epsfcheck\figins}
\def\epsfcheck{\ifx\epsfbox\UnDeFiNeD
\message{(NO epsf.tex, FIGURES WILL BE IGNORED)}
\gdef\figin##1{\vskip2in}\gdef\figins##1{\hskip.5in}% blank space instead
\else\message{(FIGURES WILL BE INCLUDED)}%
\gdef\figin##1{##1}\gdef\figins##1{##1}\fi}
\def\DefWarn#1{}
\def\figinsert{\goodbreak\midinsert}  % instead \topinsert
\def\ifig#1#2#3{\DefWarn#1\xdef#1{Fig.~\the\figno}
\writedef{#1\leftbracket fig.\noexpand~\the\figno}%
\figinsert\figin{\centerline{#3}}\medskip\centerline{\vbox{\baselineskip12pt
\advance\hsize by -1truein\noindent\footnotefont\centerline{{\bf
Fig.~\the\figno}\ \sl #2}}}
\bigskip\endinsert\global\advance\figno by1}

%%%%%%%%  Second line in Figure caption
\def\iifig#1#2#3#4{\DefWarn#1\xdef#1{Fig.~\the\figno}
\writedef{#1\leftbracket fig.\noexpand~\the\figno}%
\figinsert\figin{\centerline{#4}}\medskip\centerline{\vbox{\baselineskip12pt
\advance\hsize by -1truein\noindent\footnotefont\centerline{{\bf
Fig.~\the\figno}\ \ \sl #2}}}\smallskip\centerline{\vbox{\baselineskip12pt
\advance\hsize by -1truein\noindent\footnotefont\centerline{\ \ \ \sl #3}}}
\bigskip\endinsert\global\advance\figno by1}

%%%%%%%%%%%%%%%%%%%%%%%%%%%%%%%%%%%%%%%%%%%%%%%%%%%%%%%%%%%%%%%%%%%%%
%%%%%%%%%%%%%%%   Standard alltime definitions   %%%%%%%%%%%%%%%%%%%%
%%%%%%%%%%%%%%%%%%%%%%%%%%%%%%%%%%%%%%%%%%%%%%%%%%%%%%%%%%%%%%%%%%%%%

\def\tilde{\widetilde}

\def\h {{1\over 2}}

\def\ov {\overline}
\def\o {\over}
\def\fc#1#2{{#1 \o #2}}

\def\IZ{ {\bf Z}}\def\IP{{\bf P}}\def\IC{{\bf C}}\def\IR{ {\bf R}}\def\IN{ {\bf N}}
\def\IQ{ {\bf Q}}

      % For Eisenstein E2
  % For Polylogarithm

\def\br{\hfill\break}

\def\mod {{\rm mod}}
\def\lf {\left}
\def\ri {\right}
\def\ra {\rightarrow}
\def\lra {\longrightarrow}

\def\Lc {{\cal L}} 
\def\Mc {{\cal M}} \def\Ac {{\cal A}}
 
 \def\Uc {{\cal U}}

%%%%%%%%  My Shuffle Product
\def\shuffle{{\hskip0.10cm \vrule height 0pt width 8pt depth 0.75pt  \hskip-0.3cm\ss{\rm III}\hskip0.05cm}}
%%%%%%%%%%%%%%%%%%%%%%%%%%%%%%%%%%%%%%%%%%%%%%%%%%%%%%%%%%%%%%%%%%%
\def\sv{{\rm sv}}
\def\SVM{{\zeta^m_{\rm sv}}}
\def\SV{{\zeta_{\rm sv}}}
\def\ad{{\rm ad}}
%%%%%%%%%%%%%%%%%%%%%%%%%%%%%%%%%%%%%%%%%%%%%%%%%%%%%%%%%%%%%%%%

\lref\Deligne{
P. Deligne,
``Le groupe fondamental de la droite projective moins trois points,'' 
in: Galois groups over $\IQ$, Springer, MSRI publications {\bf 16} (1989), 72-297;
``Periods for the fundamental group,'' 
Arizona Winter School 2002.}

\lref\seealso{
H. Furusho, 
``Four groups related to associators,''
[arXiv:1108.3389 [math.QA]];
``p-adic multiple zeta values II. Tannakian interpretations,''
American Journal of Mathematics, Vol. {\bf 129} (2007), 1105-1144.
}

\lref\BroedelTTA{
J.~Broedel, O.~Schlotterer and S.~Stieberger,
 ``Polylogarithms, Multiple Zeta Values and Superstring Amplitudes,''
Fortsch.\  Phys.\  {\bf 61}, 812 (2013).
[arXiv:1304.7267 [hep-th]].
%%CITATION = DAMTP-2013-22%%
}

\lref\Ihara{
Y. Ihara, 
``Some arithmetic aspects of Galois actions on the pro-$p$ fundamental group of
$\IP\backslash\{0,1,\infty\}$,'' Proc. of Symposia in Pure Mathematics Vol. {\bf 70} (2002).
}

\lref\Iharai{
Y. Ihara, 
``On beta and gamma functions associated with the Grothendieck--Teichm\"uller group,'' 
in: Aspects of Galois Theory, London Math. Soc. Lecture Notes  {\bf 256} (1999), 144-179;
``On beta and gamma functions associated with the Grothendieck-Teichm\"uller group. II,''  
J. reine u. angew. Math. {\bf 527} (2000), 1-11;\br
H. Furusho,
``Multiple zeta values and Grothendieck-Teichm\"{u}ller groups,''  
AMS Contemporary Math., Vol. {\bf 416}, (2006), 49-82
}

\lref\SS{
  O.~Schlotterer and S.~Stieberger,
``Motivic Multiple Zeta Values and Superstring Amplitudes,''
[arXiv:1205.1516 [hep-th]], to appear in  J. Phys. A: Math. Theor.
%%CITATION = arXiv:1205.1516%%
}

\lref\SVMZV{
  F. Brown,
``Single-valued periods and multiple zeta values,''
[arXiv:1309.5309 [math.NT]].
%%CITATION = arXiv:1309.5309%%
}

\lref\BrownPoly{
F. Brown, 
``Single-valued multiple polylogarithms in one variable,''
C.R. Acad. Sci. Paris, Ser. I {\bf 338}, 527-532 (2004).}

\lref\Schnetz{
  O.~Schnetz,
``Graphical functions and single-valued multiple polylogarithms,''\br
[arXiv:1302.6445 [math.NT]].
%%CITATION = arXiv:1302.6445%%
}

\lref\StiebergerHZA{
  S.~Stieberger and T.R.~Taylor,
``Superstring Amplitudes as a Mellin Transform of Supergravity,''
Nucl.\ Phys.\ B {\bf 873}, 65 (2013) 
[arXiv:1303.1532 [hep-th]];
%%CITATION = MPP--2013--18%%
``Superstring/Supergravity Mellin Correspondence in Grassmannian Formulation,''
Phys.\ Lett.\ B {\bf 725}, 180 (2013)  
[arXiv:1306.1844 [hep-th]].
%%CITATION = MPP-2013-148%%
}

\lref\BjerrumBohrRD{
  N.E.J.~Bjerrum-Bohr, P.H.~Damgaard and P.~Vanhove,
  ``Minimal Basis for Gauge Theory Amplitudes,''
  Phys.\ Rev.\ Lett.\  {\bf 103}, 161602 (2009)
  [arXiv:0907.1425 [hep-th]].
  %%CITATION = PRLTA,103,161602;%%
}

\lref\StiebergerHQ{
  S.~Stieberger,
  ``Open \& Closed vs. Pure Open String Disk Amplitudes,''
  arXiv:0907.2211 [hep-th].
  %%CITATION = ARXIV:0907.2211;%%
}

\lref\BSST{J. Broedel, O. Schlotterer, S. Stieberger, and T. Terasoma, 
work in progress.}

\lref\Brown{
  F.~Brown,
``On the decomposition of motivic multiple zeta values,''
 in `Galois-Teichm\"uller Theory and Arithmetic Geometry', Advanced Studies in Pure Mathematics 63 (2012) 31-58 [arXiv:1102.1310 [math.NT]]. 
%%CITATION = arXiv:1102.1310%%
}

\lref\Goncharov{
A.B. Goncharov, 
``Galois symmetries of fundamental groupoids and noncommutative geometry,''
 Duke Math. J. 128 (2005) 209-284. [arXiv:math/0208144v4 [math.AG]].
}

\lref\GRAV{
  S.~Stieberger,
 ``Constraints on Tree-Level Higher Order Gravitational Couplings in Superstring Theory,''
Phys.\ Rev.\ Lett.\  {\bf 106}, 111601 (2011) 
[arXiv:0910.0180 [hep-th]].
%%CITATION = arXiv:0910.0180%%
}

\lref\DataMine{
J.~Bl\"umlein, D.J.~Broadhurst and J.A.M.~Vermaseren,
``The Multiple Zeta Value Data Mine,''
Comput.\ Phys.\ Commun.\  {\bf 181}, 582 (2010).
[arXiv:0907.2557 [math-ph]].
%%CITATION = arXiv:0907.2557%%
}

\lref\KawaiXQ{
  H.~Kawai, D.C.~Lewellen and S.H.H.~Tye,
``A Relation Between Tree Amplitudes Of Closed And Open Strings,''
  Nucl.\ Phys.\  B {\bf 269}, 1 (1986).
  %%CITATION = NUPHA,B269,1;%%
}

\lref\GoldenXVA{
  J.~Golden, A.B.~Goncharov, M.~Spradlin, C.~Vergu and A.~Volovich,
``Motivic Amplitudes and Cluster Coordinates,''
[arXiv:1305.1617 [hep-th]].
%%CITATION = arXiv:1305.1617%%
}

\lref\BCJ{
Z.~Bern, J.J.M.~Carrasco and H.~Johansson,
  ``New Relations for Gauge-Theory Amplitudes,''
  Phys.\ Rev.\  D {\bf 78}, 085011 (2008)
  [arXiv:0805.3993 [hep-ph]];
  %%CITATION = PHRVA,D78,085011;%%
``Perturbative Quantum Gravity as a Double Copy of Gauge Theory,''
Phys.\ Rev.\ Lett.\  {\bf 105}, 061602 (2010) 
[arXiv:1004.0476 [hep-th]];\br
%%CITATION = arXiv:1004.0476%%
Z.~Bern, T.~Dennen, Y.-t.~Huang and M.~Kiermaier,
``Gravity as the Square of Gauge Theory,''
Phys.\ Rev.\ D {\bf 82}, 065003 (2010) 
[arXiv:1004.0693 [hep-th]].
%%CITATION = arXiv:1004.0693%%
}

\lref\CachazoIEA{
  F.~Cachazo, S.~He and E.Y.~Yuan,
``Scattering Equations and KLT Orthogonality,''
[arXiv:1306.6575 [hep-th]];
%%CITATION = arXiv:1306.6575%%
``Scattering of Massless Particles in Arbitrary Dimension,''
[arXiv:1307.2199 [hep-th]];
%%CITATION = arXiv:1307.2199%%
``Scattering of Massless Particles: Scalars, Gluons and Gravitons,''
[arXiv:1309.0885 [hep-th]];\br
%%CITATION = arXiv:1309.0885%%
S.~Litsey and J.~Stankowicz,
``Kinematic Numerators and a Double-Copy Formula for N = 4 Super-Yang-Mills Residues,''
[arXiv:1309.7681 [hep-th]].
%%CITATION = arXiv:1309.7681%%
}

\lref\Duhr{
 L.J.~Dixon, C.~Duhr and J.~Pennington,
 ``Single-valued harmonic polylogarithms and the multi-Regge limit,''
JHEP {\bf 1210}, 074 (2012).
[arXiv:1207.0186 [hep-th]];\br
%%CITATION = arXiv:1207.0186%%
F.~Chavez and C.~Duhr,
``Three-mass triangle integrals and single-valued polylogarithms,''
JHEP {\bf 1211}, 114 (2012).
[arXiv:1209.2722 [hep-ph]];\br
%%CITATION = arXiv:1209.2722%%
V.~Del Duca, L.J.~Dixon, C.~Duhr and J.~Pennington,
``The BFKL equation, Mueller-Navelet jets and single-valued harmonic polylogarithms,''
[arXiv:1309.6647 [hep-ph]].
%%CITATION = arXiv:1309.6647%%
}

\lref\LeurentMR{
  S.~Leurent and D.~Volin,
``Multiple zeta functions and double wrapping in planar $N=4$ SYM,''
Nucl.\ Phys.\ B {\bf 875}, 757 (2013).
[arXiv:1302.1135 [hep-th]].
%%CITATION = NORDITA-2013-11%%
}

\lref\Drummond{
  J.M.~Drummond and E.~Ragoucy,
``Superstring amplitudes and the associator,''
JHEP {\bf 1308}, 135 (2013).
[arXiv:1301.0794 [hep-th]].
%%CITATION = arXiv:1301.0794%%
}

\lref\Broedel{
  J.~Broedel, O.~Schlotterer, S.~Stieberger and T.~Terasoma,
``All order alpha'-expansion of superstring trees from the Drinfeld associator,''
[arXiv:1304.7304 [hep-th]].
%%CITATION = DAMTP-2013-23%%
}

\lref\KnizhnikNR{
  V.G. Knizhnik and A.B. Zamolodchikov,
``Current Algebra and Wess-Zumino Model in Two-Dimensions,''
Nucl.\ Phys.\ B {\bf 247}, 83 (1984).
}

\lref\Drini{
  V.G. Drinfeld,
``Quasi Hopf algebras,''
Alg. Anal. {\bf 1}, 114 (1989); 
English translation: Leningrad Math. J. {\bf 1} (1989), 1419-1457.
}

\lref\Drinii{
V.G. Drinfeld, 
``On quasitriangular quasi-Hopf algebras and on a group that is closely connected with 
$Gal(\ov\IQ/\IQ)$,'' 
Alg. Anal. {\bf 2}, 149 (1990);
English translation: Leningrad Math. J. {\bf 2} (1991), 829-860.
}

\lref\LeMurakami{
T.Q.T. Le and J. Murakami,
``Kontsevich's integral for the Kauffman polynomial,''
Nagoya Math. J. {\bf 142} (1996), 39-65.
}

\lref\ElvangCUA{
  H.~Elvang and Y.-t.~Huang,
``Scattering Amplitudes,''
[arXiv:1308.1697 [hep-th]].
%%CITATION = arXiv:1308.1697%%
}

\lref\MafraNV{
  C.R.~Mafra, O.~Schlotterer and S.~Stieberger,
``Complete N-Point Superstring Disk Amplitude I. Pure Spinor Computation,''
Nucl.\ Phys.\ B {\bf 873}, 419 (2013).
[arXiv:1106.2645 [hep-th]];
%%CITATION = arXiv:1106.2645%%
``Complete N-Point Superstring Disk Amplitude II. Amplitude and Hypergeometric Function Structure,''
Nucl.\ Phys.\ B {\bf 873}, 461 (2013).
[arXiv:1106.2646 [hep-th]].
%%CITATION = arXiv:1106.2646%%
}

%%%%%%%%%%%%%%%%%%%%%%%%%%%%%%%%%%%%%%%%%%%%%%%%%%%%%%%%%%%%%%%%%%%

\Title{\vbox{\rightline{MPP--2013--278}
}}
{\vbox{\centerline{Closed Superstring Amplitudes, Single--Valued}\br
\centerline{ Multiple Zeta Values and Deligne Associator}}}
\medskip
\centerline{S. Stieberger}
\bigskip
\centerline{\it Max--Planck--Institut f\"ur Physik}
\centerline{\it Werner--Heisenberg--Institut, 80805 M\"unchen, Germany}

\vskip15pt

\medskip
\bigskip\bigskip\bigskip
\centerline{\bf Abstract}
\vskip .2in
\noindent

We revisit the tree--level closed superstring amplitude  and identify
its $\ap$--expansion as series with  single--valued multiple zeta values as coefficients.
The latter represent  a subclass of multiple 
zeta values originating from single--valued multiple polylogarithms at unity.
Moreover, the $\ap$--expansion of the closed superstring amplitude 
can be cast into the same algebraic form as the open superstring amplitude:
the closed superstring amplitude essentially is the single--valued version of the open superstring amplitude. This fact points into a deeper connection between gauge and gravity amplitudes than what is implied by Kawai--Lewellen--Tye relations. 
Furthermore, we argue, that the Deligne associator carries the relevant information on the
closed superstring amplitude. In particular,  we give an explicit representation of the Deligne associator in terms of Gamma functions
modulo squares of  commutators of the underlying Lie algebra. This form of the associator can be
interpreted as the four--point closed superstring amplitude.

\Date{}
\noindent
\goodbreak
%\listtoc 
%\writetoc
\break
%%%%%%%%%%%%%%%%%%%%%%%%%%%%%%%%%%%%%%%%%%%%%%%%%%%%%%%%%%%%%%%%%%%%%%%%%%%%%%%
\newsec{Introduction}

During the last years a great deal of work has been addressed to the problem of revealing and understanding the hidden mathematical structures of scattering amplitudes in both field-- and string theory, for a recent review \cf \ElvangCUA.
Particular emphasis on the underlying algebraic structure of amplitudes seems to be especially 
fruitful and might eventually yield an alternative way of constructing perturbative amplitudes
by methods residing in arithmetic algebraic geometry. In particular, studying motivic aspects of amplitudes has dramatically changed our view of how to write amplitudes in terms of simple 
objects, {\it cf.}~\Goncharov\ for an early and \GoldenXVA\ for a recent reference.  
Although motivic amplitudes seem to be mathematically more complicated, they are much more structured, organized and canonical objects.

In perturbative string theory, it is the dependence on the inverse string tension $\ap$, \ie the nature of the underlying string world--sheet describing the string interactions, which provides an extensive and rich structure in the analytic expressions  of the amplitudes. Some of the motivic concepts have recently matured in describing tree--level superstring amplitudes \SS.
By passing from the multiple zeta values (MZVs) entering as coefficients in the $\ap$--expansion of the amplitude to their motivic  versions \refs{\Goncharov,\Brown} 
and then mapping the latter to elements of a Hopf algebra 
reveals the motivic structure of the superstring amplitude. In this way the motivic superstring amplitude becomes a rather simple and well organized object. At the same time it is completely insensitive to a change of the basis of the underlying MZVs.

Perturbative gauge and gravity amplitudes in string theory seem to be rather different due to the 
unequal world--sheet topology of open and closed strings.
Although in practice some properties of scattering amplitudes in both gauge and gravity theories suggest a deeper relation originating from string theory, it is not clear how and whether more
symmetries or analogies between open and closed string amplitudes can be found. 
Finding more connections between open and closed string amplitudes is one aim of this article.

In {\it Ref.} \SS\ the closed superstring tree--level amplitude has been presented and it has been observed, that in contrast to the open 
string case only  MZVs of a special class show up in its  $\ap$--expansion.
In Section 2 we revisit the tree--level closed superstring amplitude  and identify the coefficients of its power series in~$\ap$ as single--valued multiple zeta values (SVMZVs).
The latter  represent a subclass of MZVs originating from single--valued multiple polylogarithms (SVMPs) at unity \doubref\BrownPoly\Schnetz. We find, that the $\ap$--expansion of the closed superstring amplitude can be obtained from that of the open superstring amplitude
by simply replacing MZVs by their corresponding SVMZVs.
In his recent work \SVMZV\  Brown has introduced the
 map $\sv$, which maps the algebra of non--commutative words describing the open superstring amplitude to a smaller subalgebra, which describes the space of SVMZVs. In Section 3 we find, that
the closed superstring amplitude essentially follows from the open superstring amplitude
by applying this map~$\sv$. 
The Drinfeld associator \doubref\Drini\Drinii, which is an infinite series in two non--commutative variables with coefficients being MZVs, has been argued to be the generating function of the open superstring amplitudes \doubref\Drummond\Broedel.
In Section~4 we identify the Deligne associator \Deligne, 
which has SVMZVs as coefficients in its series, 
to be the relevant object describing closed superstring amplitudes. Furthermore, we give an explicit representation of the Deligne associator in terms of Gamma functions
modulo squares of  commutators of the underlying Lie algebra and this form can be
interpreted as the four--point closed superstring amplitude.
Finally, in Section 5 we give some concluding remarks.

\newsec{Closed superstring amplitudes and single--valued multiple zeta values}

In this Section we want to illuminate the observations on the $\ap$--expansion of the graviton amplitude \SS\ in view of Browns recent work on SVMZVs \SVMZV. 
We shall find a striking similarity between the open superstring amplitude~$\Ac$ and 
the closed superstring amplitude~$\Mc$ thus giving rise to a new relation between gauge and 
gravity amplitudes.

The string world--sheet describing the tree--level string $S$--matrix of $N$ 
gravitons has the topology of a complex sphere with $N$ insertions
of graviton vertex operators. Of the latter $N-3$ are integrated on the whole
sphere leading to the following type of complex integrals
\eqn\type{
 \lf(\prod_{j=2}^{N-2}\int\limits_{z_j\in\IC} d^2z_j\ri) \prod_{1\leq i<j\leq N-1}
|z_i-z_j|^{s_{ij}}\ (z_j-z_i)^{n_{ij}}\ ,}
with $z_1=0,\ z_{N-1}=1,\ z_N=\infty$, the set of integers $n_{ij}\in\IZ$ and the real numbers 
$s_{ij}=\ap (k_i+k_j)^2=2\ap k_ik_j$. 
The latter describe the $\h N(N-3)$ independent kinematic invariants of the scattering process
involving $N$ external momenta $k_i,\ i=1,\ldots,N$ and $\ap$ is the inverse string tension. 
The integrals \type\ can be considered as iterated integrals on $\IP^1\backslash\{0,1,\infty\}$ 
integrated independently on all choices of paths.

One of the key properties of graviton amplitudes in string theory is that 
at tree--level they can be expressed as sum over squares of (color ordered) 
gauge amplitudes in the left-- and right--moving sectors. 
This map, known as Kawai--Lewellen--Tye (KLT) relations  \KawaiXQ, 
gives a relation between a closed string tree--level amplitude $\Mc$ involving $N$ closed strings and a sum of squares of (partial ordered) open string tree--level amplitudes.  
We may write these relations in matrix notation as follows\eqn\graviton{
\Mc(1,\ldots,N)=\Ac^t\ S\ \Ac\ ,}
with the vector $\Ac$ encoding a basis of $(N-3)!$ open string subamplitudes 
and some $(N-3)!\times(N-3)!$ intersection matrix $S$. 
The KLT relations  are insensitive
to the compactification details and the amount of supersymmetries of the superstring background. 
Hence, the following discussions and results on $N$--graviton tree--level scattering are completely general. Since open superstring amplitudes $\Ac$ are needed to describe the closed superstring amplitude $\Mc$, in the following let us first review some aspects of open superstring amplitudes.

Tree--level scattering of $N$ open strings involves $(N-3)!$ independent
color ordered subamplitudes \doubref\BjerrumBohrRD\StiebergerHQ. The latter can be collected in an
$(N-3)!$--dimensional vector $\Ac$, which can be expressed as \doubref\MafraNV\SS
\eqn\canbewritten{
\Ac=F\ A\ ,}
with the $(N-3)!$--dimensional vector $A$ encoding the Yang--Mills basis and the period matrix 
$F$, given by\foot{The ordering colons 
$:\ldots :$ are defined such that matrices with larger subscript multiply matrices with smaller subscript from the left, \ie $: \, M_{i} \ M_{j} \, : =  
\cases{M_{i} \ M_j\ , & $i \geq j\ ,$\cr 
       M_{j} \ M_i\ , &              $i<j$\ .}$
The generalization to iterated matrix products $: M_{i_1} M_{i_2} \ldots M_{i_p}:$ is straightforward.} \SS
\eqn\period{
F=P\ Q\ :\exp\lf\{\sum\limits_{n\geq1}\zeta_{2n+1}\ M_{2n+1}\ri\}:\ ,}
with the $(N-3)!\times (N-3)!$ matrices
\eqn\setdef{\eqalign{
P&=1+\sum_{n\geq 1} \z_2^n\ P_{2n}\ \ \ ,\ \ \ P_{2n}=\lf.F\ri|_{\zeta_2^n}\ ,\cr
M_{2n+1}&=\lf.F\ri|_{\zeta_{2n+1}}\ ,}}
and: 
\eqn\QQ{\eqalign{
Q:=1+\sum_{n\geq 8}Q_n&=1+\fc{1}{5}\ \zeta_{3,5}\ [M_5,M_3]+\lf\{\ \fc{3}{14}\ \zeta_5^2+\fc{1}{14}\ \zeta_{3,7}\ \ri\}\ [M_7,M_3]\cr
&+\lf\{\ 9\ \zeta_2\ \zeta_9+\fc{6}{25}\ \zeta_2^2\ \zeta_7-\fc{4}{35}\ \zeta_2^3\ \zeta_5+\fc{1}{5}\ \zeta_{3,3,5}\ \ri\}\ [M_3,[M_5,M_3]]+\ldots\ .}}
Above we have have adapted to the following definition of MZVs
\eqn\MZV{
\zeta_{n_1,\ldots,n_r}:=\zeta(n_1,\ldots,n_r)=
\sum\limits_{0<k_1<\ldots<k_r}\ \prod\limits_{l=1}^r k_l^{-n_l}\ \ \ ,\ \ \ n_l\in\IN^+\ ,\ n_r\geq2\ ,}
with $r$ specifying the depth and $w=\sum_{l=1}^rn_l$ denoting 
the weight of the MZV $\zeta_{n_1,\ldots,n_r}$. 
Furthermore,  we have used the MZV basis constructed in \DataMine.
Note, that for any $N$ the tree--level open superstring amplitude assumes the form \canbewritten\
with \period. The only ingredients are the $(N-3)!\times (N-3)!$ matrices $P_{2n}$ and $M_{2n+1}$, whose entries are polynomials in degree $2n$ and $2n+1$ in the kinematic invariants $s_{ij}$, respectively. For different $N$ the matrices $P_{2n}$ and $M_{2n+1}$ have been thoroughly  investigated in \BroedelTTA.
Moreover,  the form of the expressions \period, \setdef\ and \QQ\ is bolstered by the algebraic structure of motivic MZVs and their decomposition \Brown.
In fact, the operator $F$ is isomorphic to the  decomposition operator of motivic MZVs~\SS.

Applying the open string results \period\ to the graviton amplitude \graviton\ gives rise to \SS:
\eqn\Graviton{
\Mc=A^t\ G\ A\ ,}
with the matrix\foot{Note, that the transpositions involved in the expression $\lf( : \exp\lf\{\sum_{r}  \zeta_{r}\ M_{r}^t \ri\} : \ri)^t$ lead to a reversal of the matrix multiplication order compared to the ordered product $:\exp\lf\{\sum_{s}  \zeta_{s}\ M_{s}\ri\} :$ without transposition, \ie:
$\lf( : \exp\lf\{\sum_{r\in 2\IN^+ +1}   \zeta_{r}\ M_{r}^t \ri\} : \ri)^t= 1+\zeta_3 M_3 + \zeta_5 M_5 + {1 \over 2} \zeta_3^2 M_3^2 + \zeta_7 M_7 + \zeta_3 \zeta_5 M_3 M_5  
+{1 \over 6} \zeta_3^3 M_3^3+ \zeta_9 M_9 
 + {1 \over 2} \zeta_5^2 M_5^2 + \zeta_3 \zeta_7 M_3 M_7 + {1 \over 2} \zeta_3^2 \zeta_5 M_3^2 M_5 + \zeta_{11} M_{11} + \ldots \ .$}
\eqn\Gmatrix{
G=F^t\ S\ F=
S_0\ \lf(  \exp\lf\{\sum_{r\in 2\IN^+ +1} \! \! \!  \zeta_{r}\ M_{r}^t \ri\}  \ri)^t \tilde Q\ Q\   \exp\lf\{\sum_{s \in 2\IN^++1} \! \! \!  \zeta_{s}\ M_{s}\ri\}\ ,} 
and the intersection form $S_0$ defined by:
\eqn\intersection{
S_0=P^t\ S\ P\ .}
An other interpretation of $S_0$ is, that it makes sure, that the field--theory limit
of the graviton amplitude \Graviton\ is correctly reproduced: $\lf.\Mc(1,\ldots,N)\ri|_{\ap\ra 0}=
A^t\ S_0\ A$, \ie $\lf.G\ri|_{\ap\ra 0}=S_0$.

It has already been observed in \SS\ (extending the results \GRAV), that one implication of the specific form of \Gmatrix\ is, that only a certain  subclass of MZVs appears in the 
$\ap$--expansion of the graviton amplitude \Graviton.
In fact, in \eqq  \Gmatrix\ the product $\tilde Q Q$ is given by~\SS 
\eqn\freeeven{
\tilde Q\ Q = 1 + 2\ Q_{11} + 2\ Q_{13}+2\ Q_{15}+\ldots \ ,}
with $\tilde Q = \lf.Q\ \ri|_{\Qc_{(r)} \rightarrow (-1)^{r+1}\Qc_{(r)}}$ and $\Qc_{(r)}$
any nested commutator of depth $r$ appearing in \QQ.
As a consequence the product \freeeven\ is free of odd powers in even depth commutators 
$\Qc_{(2n)}$.
Furthermore, the specific form of \Gmatrix\ involves, that  MZVs of even weight or depth $\geq 2$ only enter through the product \freeeven\ starting at weight $w=11$ \SS.

The subclass of MZVs appearing in \Gmatrix\ can be identified
as single--valued multiple zeta values 
\eqn\trivial{
\SV(n_1,\ldots,n_r)\in\IR}
originating from single--valued multiple polylogarithms at unity and studied recently in \SVMZV\ from a mathematical point of view. 
Let us now illuminate \Gmatrix\ in view of this subclass of MZVs.
The numbers \trivial\ satisfy the same double shuffle and associator relations than the 
usual MZVs and many more relations \SVMZV:
\eqn\Example{\eqalign{
\SV(2)&=0\ ,\cr
\SV(2n+1)&=2\ \zeta_{2n+1}\ ,\ \ \ n\geq 1\ .}}
Furthermore, for instance we have computed:
\eqnn\Examplee{
$$\eqalignno{
\SV(3,5)&=-10\ \z_3\ \z_5\ ,\cr
%%  \SV(5,3)&=14\ \z_3\ \z_5\ ,\cr
\SV(3,7)&=-28\ \z_3\ \z_7-12\ \z_5^2\ ,\cr
\SV(3,3,5)&=2\ \z_{3,3,5}-5\ \z_3^2\ \z_5+90\ \z_2\ \z_9+\fc{12}{5}\ \z_2^2\ \z_7-\fc{8}{7}\ \z_2^3\ \z_5^2\ ,\cr 
\SV(3,9)&=-54\ \z_3\ \z_9-42\ \z_5\ \z_7\ ,\cr
\SV(5,7)&=-84\ \z_3\ \z_9-63\ \z_5\ \z_7\ ,&\Examplee\cr
\SV(1,1,4,6)&=\fc{6397}{36}\ \z_3\ \z_9+\fc{1597}{8}\ \z_5\ \z_7+\fc{4}{3}\ \z_3^4\ ,\cr
%%  \SV(3,3,5)&=-\z_{3,5,3}+\fc{299}{2}\ \z_{11}+\z_3\ \z_{3,5}-5\ \z_3^2\ \z_5\ ,\cr 
%%  \SV(3,5,3)&=2\ \z_{3,5,3}-2\ \z_3\ \z_{3,5}-10\ \z_3^2\ \z_5\ ,\cr
\SV(3,5,5)&=2\ \z_{3,5,5}+10\ \z_5\ \z_{3,5}+50\ \z_3\ \z_5^2+275\ \z_2\ \z_{11}+20\ \z_2^2\ 
\z_9\ ,\cr
\SV(3,3,7)&=2\ \z_{3,3,7}+12\ \z_5 \z_{3,5}-14\ \z_3^2 \z_7+60\ \z_3 \z_5^2+407\ \z_2 \z_{11}+\fc{112}{5}\ \z_2^2 \z_9-\fc{64}{35}\ \z_2^3 \z_7.}$$}

The matrix \Gmatrix\ can be written purely in terms of SVMZVs \trivial\ as follows:
\eqnn\gravlowmom{
$$\eqalignno{
G &=  
 S_0\ \big( \ 1+\SV(3)\ M_3 + \SV(5)\ M_5 + \h\ \SV(3)^2\ M_3^2 + \SV(7)\ M_7 \cr
& +\h\ \SV(3)\ \SV(5)\ \ \{M_3, M_5\} + \SV(9)\ M_9 + {1 \over 3!}\ \SV(3)^3\ M_3^3 
 \cr 
& + \h\ \SV(5)^2\ M_5^2  + \h\ \SV(3)\ \SV(7)\ \{M_3 ,M_7\} \cr
&  + Q_{\rm sv}(11)+ \SV(11)\ M_{11} + \fc{1}{8}\ \SV(3)^2\ \SV(5)\ \{ M_3,\{M_3,M_5\}\} \cr
&  + {1 \over 4!}\ \SV(3)^4\ M_3^4+\h\ \SV(3)\ \SV(9)\ \{M_3, M_9 \}\cr
&+ \h\ \SV(5)\ \SV(7)\ \{M_5, M_7\} + Q_{\rm sv}(13) + \SV(13)\ M_{13} &\gravlowmom \cr
 &+ \fc{1}{8}\ \SV(3)^2\ \SV(7)\ \{ M_3,\{M_3, M_7\}\} 
 + \fc{1}{4}\ \SV(3)\ \SV(5)^2\ \{M_3, M_5^2\}\cr
 &+\h\ \SV(3)\ \{M_3 ,Q_{\rm sv}(11)\}+ \h\ \SV(7)^2\ M_7^2
  + \h\ \SV(3)\ \SV(11)\ \{M_3, M_{11}\}\cr
  & + \h\ \SV(5)\ \SV(9)\ \{M_5 ,M_9\} 
  +{1\over48}\  \SV(3)^3\ \SV(5)\ \{M_3,\{M_3,\{M_3,M_5\}\}\} + \ldots \ \big)\ ,}$$}
with the Poisson bracket $\{A,B\}=AB+BA$ and:
\eqnn\withQ{
$$\eqalignno{
Q_{\rm sv}(11)&=\lf\{\ \fc{1}{5}\ \SV(3,3,5)+\fc{1}{8}\ \SV(3)^2\ \SV(5)\ \ri\}\ 
[M_3,[M_5,M_3]]\ ,\cr
Q_{\rm sv}(13)&=\lf\{\ \fc{1}{25}\ \SV(3,5,5)-\fc{1}{4}\ \SV(3)\ \SV(5)^2\ \ri\}\ 
[M_5,[M_5,M_3]]&\withQ\cr
&+\lf\{\fc{1}{14}\ \SV(3,3,7)-\fc{3}{35}\ \SV(3,5,5)+\fc{1}{8}\ \SV(3)^2\ \SV(7)\ri\}\ [M_3,[M_7,M_3]]\ .}$$} 
Note, that in \gravlowmom\ the terms \freeeven\ containing  MZVs of even weight or 
depth $\geq 2$ comprise into the expressions $Q_{\rm sv}(n)$, which can be written purely in terms of SVMZVs. 

In fact, we can go one step further by introducing the following homomorphism:
\eqn\mapSV{
\sv: \z_{n_1,\ldots,n_r}\mapsto\ \SV( n_1,\ldots,n_r )\ .}
With \Example\ we have:
\eqn\Example{\eqalign{
\sv(\z_2)&=\SV(2)=0\ ,\cr
\sv(\z_{2n+1})&=\SV(2n+1)=2\ \zeta_{2n+1}\ ,\ \ \ n\geq 1\ .}}
We can apply this map \mapSV\ to the period matrix $F$ given in \period
\eqn\applyf{
\sv(F)=sv(Q)\ :\exp\lf\{\sum\limits_{n\geq1}\SV(2n+1)\ M_{2n+1}\ri\}:\ ,}
and obtain:
\eqnn\applyF{
$$\eqalignno{
\sv(F)&=1+\SV(3)\ M_3+\SV(5)\ M_5+\h\ \SV(3)^2\ M_3^2+\SV(7)\ M_7\cr
&+\SV(3)\ \SV(5)\ M_5\ M_3+\fc{1}{5}\ \SV(3,5)\ [M_5,M_3]+\SV(9)\ M_9
+\fc{1}{3!}\ \SV(3)^3\ M_3^3\cr
&+\SV(3)\ \SV(7)\ M_7\ M_3+\h\ \SV(5)^2\ M_5^2+\lf(\fc{3}{14}\SV(5)^2+\fc{1}{14}\SV(3,7)\ri)\ [M_7,M_3]\cr
&+\fc{1}{5}\ \SV(3,3,5)\ [M_3,[M_5,M_3]]+\fc{1}{5}\ \SV(3)\ \SV(3,5)\ [M_5,M_3]M_3\cr
&+\SV(11)\ M_{11}+\h\ \SV(3)^2\ \SV(5)\ M_5M_3^2+\ldots\ .&\applyF}$$}

\comment{
 For a given weight
$w$ we consider $\sv(F_w)$, with $F_w$ being the weight $w$ part of $F$. {\it E.g.} at weights 
$w=8,10,11$ and $w=12$ we find:}

\comment{\eqnn\applyF{
$$\eqalignno{
\sv(F_8)&=\sv\lf(\z_2^4\ P_8+\h\ \z_2\z_3^2\ P_2M_3^2+\z_3\z_5\ M_5M_3+
\fc{1}{5}\ \z_{3,5}\ [M_5,M_3]\ri)\cr
&=\SV(3)\ \SV(5)\ M_5M_3+\fc{1}{5}\ \SV(3,5)\ [M_5,M_3]=\h\ \SV(3)\ \SV(5)\ \{M_3,M_5\}\ ,\cr
\sv(F_{10})&=\lf(\fc{3}{14}\ \SV(5)^2+\fc{1}{14}\ \SV(3,7)\ri)\ [M_7,M_3]+\SV(3)\ \SV(7)\ M_7M_3+
\h\ \SV(5)^2\ M_5^2\cr
&=\h\ \SV(3)\ \SV(7)\ \{M_3,M_7\}+\h\ \SV(5)^2\ M_5^2\ ,&\applyF\cr
\sv(F_{11})&=\fc{1}{5}\ \SV(3,3,5)\ [M_3,[M_5,M_3]]+\SV(11)\ M_{11}+\h\ \SV(3)^2\ \SV(5)\ M_3M_5M_3\ ,\cr
\sv(F_{12})&=\h\ \SV(3)\ \SV(9)\ \{M_3,M_9\}+\h\ \SV(5)\ \SV(7)\ \{M_5,M_7\}
+\fc{1}{24}\ \SV(3)^4M_3^4\ .}$$}}

\comment{A different representation in terms of the SVMZV basis chosen in \SVMZV\ gives:
\eqn\withq{\eqalign{
Q_{\rm sv}(11)&=\lf\{-\fc{1}{10}\ \SV(3,5,3)-\fc{1}{8}\ \SV(3)^2\ \SV(5)+\fc{299}{20}\ \SV(11)\ri\}\ [M_3,[M_5,M_3]]\ ,\cr
Q_{\rm sv}(13)&=\lf\{-\fc{1}{50}\ \SV(5,3,5)-\fc{3}{10}\ \SV(3)\ \SV(5)^2+\fc{1003}{100}\ \SV(13)\ri\}\ [M_5,[M_5,M_3]]\cr
&+\lf\{\fc{3}{70}\ \SV(5,3,5)-\fc{1}{28}\ \SV(3,7,3)-\fc{1}{8}\ \SV(3)^2\ \SV(7)+\fc{571}{140}\ \SV(13)\ri\}[M_3,[M_7,M_3]].}}}

%%%%  In these lines we conjecture, that the operator 
%%%% $S_0^{-1}G$ describes the decomposition of motivic SVMZVs. 

\noindent
By comparing the image \applyF\ with the  expression (in brackets) in  \gravlowmom\ we find agreement.
In fact, in general we find:
\eqn\wefind{
G=S_0\ \sv(F)\ ,\ \ \ i.e.:\ \ \  F^tSF=S_0\ \sv(F)\ .}
Thus, the effect of the kernel $S$ is to project onto a constrained period matrix $\sv(F)$.
With the equality \wefind\ the graviton amplitude \Graviton\ becomes
\eqn\GRAVITON{
\Mc=A^t\ S_0\ \sv(F)\ A\ ,}
which gives rise to the following relation between  the (tree--level) open \canbewritten\ and closed \Graviton\ superstring amplitudes
\eqn\Wefind{
\Mc=A^t\ S_0\ \sv(\Ac)\ ,} 
with the intersection matrix $S_0$ defined in \intersection\ and the vector $A$ of Yang--Mills subamplitudes introduced in \canbewritten.  

To conclude, the closed superstring amplitude $\Mc$ 
is obtained from the open superstring amplitude $\Ac$ through the map \mapSV.
In these lines the open superstring amplitude \canbewritten\  is {\it the} uniform 
form for describing both the gauge and gravity amplitudes.

\newsec{Motivic open and closed superstring amplitudes}
\def\per{{\rm per}}

In this Section after presenting the Hopf algebra structure of the open and closed superstring amplitude
 we look at the connection between the $\ap$--expansion of the open and closed  superstring amplitudes at the level of their underlying Hopf algebra.

Motivic MZVs $\zeta^m$ are defined as elements of a certain algebra $\Hc=\bigoplus_{w\geq0}\Hc_w$ over $\IQ$, which is graded for the weight and equipped with 
the  period homomorphism $\per: \Hc\ra \IR$, which maps $\zeta^m_{n_1,\ldots,n_r}$ 
to $\zeta_{n_1,\ldots,n_r}$, \ie $\per(\z^m_{n_1,\ldots,n_r})=\z_{n_1,\ldots,n_r}$.
The motivic versions of the SVMZVs \trivial\ have been defined in \SVMZV\ and are denoted by 
$\SVM(n_1,\ldots,n_r)$. The latter satisfy all
motivic relations of MZVs and $\SVM(2)=0$ \SVMZV.
The motivic SVMZVs span a subalgebra $\Hc^{\rm sv}\subset \Hc$.
There exist a homomorphism $\Hc\ra\Hc^{\rm sv}$, which maps each $\zeta^m_{n_1,\ldots,n_r}$
to $\SVM(n_1,\ldots,n_r)$.

A list of generators of $\Hc^{\rm sv}_w$ up to weight $w=14$  is collected in Tables 1--2 below.

{\vbox{\ninepoint{$$
\vbox{\offinterlineskip\tabskip=0pt
\halign{\strut\vrule#
%%%%%%%%%%%%%%%%%%
&~$#$~\hfil 
&\vrule$#$ 
&~$#$~\hfil 
&\vrule$#$ 
&~$#$~\hfil 
&\vrule$#$ 
&~$#$~\hfil 
&\vrule$#$ 
&~$#$~\hfil 
&\vrule$#$ 
&~$#$~\hfil 
&\vrule$#$ 
&~$#$~\hfil 
&\vrule$#$
&~$#$~\hfil 
&\vrule$#$ 
&~$#$~\hfil 
&\vrule$#$
&~$#$~\hfil
&\vrule$#$\cr
\noalign{\hrule}
%%%%%%%%%%%%%%%%%%
& w &&2&&3 &&4&&5 &&6&& 7 &&8 && 9 && 10  &\cr
\noalign{\hrule}
&\Hc^{\rm sv}_w&&-&&\SVM(3) &&- &&\SVM(5)&&\SVM(3)^2&& \SVM(7)&&\SVM(3)\ \SVM(5) &&\SVM(9) && \SVM(5)^2 &\cr
& &&  && &&  && && && &&  &&\SVM(3)^3  && \SVM(3)\ \SVM(7) &\cr
\noalign{\hrule}
&dim(\Hc^{\rm sv}_w) &&0&&1&&0 &&1&&1  &&1 &&1 &&2 &&2   &\cr
\noalign{\hrule}}}$$
\vskip-6pt
\centerline{\noindent{\bf Table 1:}
{\sl Generators of $\Hc^{\rm sv}_w$ for $2\leq w\leq 10$.}}
\vskip10pt}}}

{\vbox{\ninepoint{$$
\vbox{\offinterlineskip\tabskip=0pt
\halign{\strut\vrule#
%%%%%%%%%%%%%%%%%%
&~$#$~\hfil 
&\vrule$#$ 
&~$#$~\hfil 
&\vrule$#$ 
&~$#$~\hfil 
&\vrule$#$
&~$#$~\hfil
&\vrule$#$
&~$#$~\hfil
&\vrule$#$\cr
\noalign{\hrule}
%%%%%%%%%%%%%%%%%%
& w && 11 && 12&& 13 &&14 &\cr
\noalign{\hrule}
&\Hc^{\rm sv}_w  && \SVM(11)&&\SVM(3)\ \SVM(9) &&\SVM(13)&&\SVM(3)\ \SVM(3,3,5) &\cr
&   &&  \SVM(3,3,5)&&\SVM(5)\ \SVM(7) &&\SVM(3,5,5) &&\SVM(3)^3\ \SVM(5)\ &\cr
&  &&\SVM(3)^2\ \SVM(5) &&\SVM(3)^4&&\SVM(3,3,7)&&\SVM(7)^2 &\cr
&  && &&  &&\SVM(3)^2\ \SVM(7)&&\SVM(3)\ \SVM(11) &\cr
&  && &&  &&\SVM(3)\ \SVM(5)^2 && \SVM(5)\ \SVM(9)&\cr
\noalign{\hrule}
&dim(\Hc^{\rm sv}_w)  &&3 &&3 &&5 &&5  &\cr
\noalign{\hrule}}}$$
\vskip-6pt
\centerline{\noindent{\bf Table 2:}
{\sl Generators of $\Hc^{\rm sv}_w$ for $11\leq w\leq 14$.}}
\vskip10pt}}}

\noindent
Note, that the generators of Tables 1--2 are those, which appear in \gravlowmom, 
subject to the period map $\per$.

To explicitly describe the structure of the algebra $\Hc$ Brown has introduced an 
auxiliary algebra $\Uc$, the (trivial) algebra--comodule \Brown:
\eqn\introU{
\Uc=\IQ\vev{f_3,f_5,\ldots}\otimes_{\IQ} \IQ[f_2]\ .}
The first factor $\Uc'=\Uc\big\slash f_2\Uc$ is a cofree Hopf--algebra on the cogenerators 
$f_{2r+1}$ in degree $2r+1\geq 3$, whose basis consists of all non--commutative words in 
the $f_{2i+1}$. The multiplication on $\Uc'$ is given by the shuffle product $\shuffle$. 
The Hopf--algebra  $\Uc'$ is the algebra of all words constructed from the alphabet
$\{f_3,f_5,f_7,\ldots\}$ and is isomorphic to the space of non--commutative polynomials in $f_{2i+1}$.
The element $f_2$ commutes with all $f_{2r+1}$. Again, there is a grading $\Uc_k$ on $\Uc$ and 
we have the non--canonical isomorphism: $\Hc\simeq \Uc$.
Furthermore, there exists a morphism $\phi$ of graded algebra--comodules
\eqn\morphism{
\phi:\ \Hc\lra\Uc\ ,}
normalized by:
\eqn\morphismres{
\phi\big(\zeta^m_n\big)=f_n\ \ \ ,\ \ \ n\geq 2\ .}
The map \morphism, which respects the shuffle multiplication rule
\eqn\ruleshuffle{
\phi(x_1x_2)=\phi(x_1)\shuffle \phi(x_2)\ \ \ ,\ \ \ x_1,x_2\in\Hc\ ,} 
sends every motivic MZV to a non--commutative polynomial in the $f_i$.

The motivic period matrix $F^m$, where all MZVs $\zeta_{n_1,\ldots,n_r}$ are replaced by 
their motivic objects $\zeta^m_{n_1,\ldots,n_r}$, has been introduced and studied in \SS. 
Furthermore, in this reference the map of $F^m$ under 
$\phi$ has been computed, with the result:
\eqn\MOTO{
\phi(F^m) =\left( \sum_{k=0}^{\infty} f_2^k\ P_{2k} \right) \lf( \sum_{p=0}^{\infty} 
\sum_{ i_1,\ldots, i_p \atop  \in 2 \IN^+ + 1}
 f_{i_1} f_{i_2}\ldots f_{i_p}\ M_{i_p} \ldots M_{i_2} M_{i_1}\ri)\ .}
Besides, in \SS\ the motivic version of $G^m$ has been introduced and an expression for
$\phi(G^m)$ has been given. We want to rewrite the latter in view of the recent work of 
Brown~\SVMZV\ and eventually find a striking similarity between $\phi(F^m)$ and  $\phi(G^m)$.

Similar to the construction of $\Uc$ in {\it Ref.} \SVMZV\ Brown has introduced a model 
$\Uc^{\rm sv}$ for $\Hc^{\rm sv}$ via the the homomorphism
\eqn\mapsv{
\sv:\Uc'\longrightarrow\Uc'\ ,} 
with
\eqn\Mapsv{
w\longmapsto \sum_{uv=w} u\shuffle \tilde v\ ,}
and $\tilde v$ being the reversal of the word $v$.  
For the image $\Uc^{\sv}$ under \mapsv\ we have the isomorphism: $\Hc^{\rm sv}\simeq \Uc^{\rm sv}$.
For instance we have \SVMZV
\eqn\expSV{\eqalign{
\sv(f_a)&=2f_a\ ,\ \ \ \sv(f_af_b)=2\ (f_af_b+f_bf_a)\ ,\cr
\sv(f_af_bf_c)&=2\ (f_af_bf_c+f_af_cf_b+f_cf_af_b+f_cf_bf_a)\ ,\cr
\sv(f_af_bf_cf_d)&=2\ (f_af_bf_cf_d+f_af_bf_df_c+f_af_df_bf_c+f_af_df_cf_b\cr
&+f_df_af_bf_c+f_df_af_cf_b+f_df_cf_af_b+f_df_cf_bf_a)\ ,}}
for the odd integers $a,b,c,d$.
Evidently, we can extend the map \mapsv\ to $\Uc$ with: 
\eqn\mapexa{
\sv(f_2)=0\ .}

Let us now return to the matrix $G$, given in \Gmatrix\ and \gravlowmom\ and compute the 
image $\phi(G^m)$. The latter has been given in \SS\ as:
\eqn\gravsimple{
\phi(G^m )= S_0 \ \lf(\ \sum_{p=0}^{\infty} \sum_{ i_1,\ldots, i_p \atop  \in 2 \IN^+ + 1}  \ 
M_{i_1}  M_{i_2} \ldots M_{i_p} 
\ \sum_{k=0}^p  \ f_{i_1} f_{i_2} \ldots f_{i_k}  \shuffle  f_{i_p} f_{i_{p-1}} \ldots f_{i_{k+1}}\ 
\ri)\ .} 
By making profit out of the map \mapsv\ and using relations like \expSV\ we can cast \gravsimple\ into the compact form:
\eqn\MOTC{
\phi(G^m)=S_0 \ \lf(\ \sum_{p=0}^{\infty} 
\sum_{ i_1,\ldots, i_p \atop  \in 2 \IN^+ + 1} \ M_{i_1}  M_{i_2} \ldots M_{i_p} 
\ \ \sv(f_{i_1} f_{i_2}\ldots f_{i_p}) \ \ri)\ .}
After comparing \MOTC\ with \MOTO\ and using \mapexa\ we find:
\eqn\RESULT{
\phi(G^m)=S_0\ \sv(\phi(F^m))\ .}
Finally, due to \RESULT\ the motivic open superstring amplitude $\Ac^m$
\eqn\motO{
\phi(\Ac^m)=\left( \sum_{k=0}^{\infty} f_2^k\ P_{2k} \right) \lf( \sum_{p=0}^{\infty} 
\sum_{ i_1,\ldots, i_p \atop  \in 2 \IN^+ + 1}
 f_{i_1}\ldots f_{i_p}\ M_{i_p} \ldots M_{i_1}\ \  \ri)\ A\ ,}
and the motivic closed superstring amplitude $\Mc^m$
\eqn\motC{
\phi(\Mc^m)=A^t\ S_0 \ \lf(\ \sum_{p=0}^{\infty} 
\sum_{ i_1,\ldots, i_p \atop  \in 2 \IN^+ + 1} \ M_{i_1} \ldots M_{i_p} 
\ \ \sv(f_{i_1} \ldots f_{i_p}) \ \ri)\ A\ ,}
respectively, can be related as follows:
\eqn\RESULTT{
\phi(\Mc^m)=A^t\ S_0\ \sv(\phi(\Ac^m))\ ,}
with the intersection matrix $S_0$ defined in \intersection\ and the vector $A$ of Yang--Mills subamplitudes introduced in \canbewritten.
{\it Eqs.} \RESULT\ and \RESULTT\ represent the corresponding relations \wefind\ and \Wefind\
in terms of the Hopf algebra \introU, respectively. 
Note, that the map~$\phi$ can be inverted. Hence, not any information on the motivic
amplitudes $\Ac^m$ or $\Mc^m$ is lost by considering the objects $\phi(\Ac^m)$ and  $\phi(\Mc^m)$
and the full (motivic) superstring amplitudes \canbewritten\ and \Graviton\ 
can be recovered from the images $\phi(\Ac^m)$ and  $\phi(\Mc^m)$, respectively.

To conclude, the image \MOTO\ of the motivic period matrix $F^m$ under $\phi$ is {\it the} uniform 
form for both the open and closed superstring amplitude.

\newsec{Deligne associator and closed superstring amplitudes}

In his recent work \SVMZV\ Brown has identified SVMZVs as elements of the Deligne associator 
\Deligne, \ie the coefficients of the latter are the values of SVMPs at one. 
In this Section we argue, that the Deligne associator carries the relevant information on the
closed superstring amplitude and allows to extract the form of the $N$--point closed superstring amplitude. Furthermore, we give an explicit representation of the Deligne associator in terms of Gamma functions modulo squares of commutators of the underlying Lie algebra. 
This form of the associator can be interpreted as the four--point closed superstring amplitude.

The unique solution to the KZ equation\foot{Partial differential equations based on Lie algebras appear in the context of conformal field theory.} \KnizhnikNR\ 
\eqn\KZ{
\fc{d}{dz}\ L_{e_0,e_1}(z)=L_{e_0,e_1}(z)\ \lf(\fc{e_0}{z}+\fc{e_1}{1-z}\ri)\ ,}
with the generators $e_0$ and $e_1$ of the free Lie algebra $g$ can be be given as generating series of multiple polylogarithms as \BrownPoly
\eqn\genL{
L_{e_0,e_1}(z)=\sum_{w\in\{e_0,e_1\}^\times}L_w(z)\ w\ ,}
with the symbol $w\in\{e_0,e_1\}^\times$ denoting a non--commutative word 
$w_1w_2\ldots$ in the letters $w_i\in\{e_0,e_1\}$.

The generating series of SVMPs is defined by \BrownPoly
\eqn\genLc{
\Lc_{e_0,e_1}(z)=L_{-e_0,-e_1'}(\ov z)^{-1}\ L_{e_0,e_1}(z)\ ,}
% where $\tilde w$ denotes reversal of the word $w$ and 
where $e_1'$ is determined recursively by the following fixed--point equation
\eqn\fixed{
Z(-e_0,-e_1')\ e_1'\ Z(-e_0,-e_1')^{-1}=Z(e_0,e_1)\ e_1\ Z(e_0,e_1)^{-1}\ ,}
with the  Drinfeld associator $L_{e_0,e_1}(1)\equiv Z(e_0,e_1)$. The latter is given by the non--commutative generating series 
of (shuffle-regularized) MZVs \LeMurakami
\eqnn\Drinfeld{
$$\eqalignno{
Z(e_0,e_1)&=\sum_{w\in\{e_0,e_1\}^\times}\z(w)\ w=1+\z_2\ [e_0,e_1]+\z_3\ (\ [e_0,[e_0,e_1]]-[e_1,[e_0,e_1]\ )\cr
&+\z_4\ \Big(\ [e_0,[e_0,[e_0,e_1]]]-\fc{1}{4}\ [e_0,[e_1,[e_0,e_1]]]+[e_1,[e_1,[e_0,e_1]]]
+\fc{5}{4}\ [e_0,e_1]^2\ \Big)\cr
&+\z_2\ \z_3\ \Big(\ [e_0,e_1]\ (\ [e_0,[e_0,e_1]]-[e_1,[e_0,e_1]\ )-[e_0,[e_1,[e_0,[e_0,e_1]]]]\cr
&+[e_0,[e_1,[e_1,[e_0,e_1]]]]\ \Big)+\z_5\ \Big(\ [e_0,[e_0,[e_0,[e_0,e_1]]]]
-\h\ [e_0,[e_0,[e_1,[e_0,e_1]]]]\cr
&-\fc{3}{2}\ [e_1,[e_0,[e_0,[e_0,e_1]]]]
+(e_0\leftrightarrow e_1)\ \Big)+\ldots\ ,&\Drinfeld}$$}
with $\z(e_1e_0^{n_1-1}\ldots e_1e_0^{n_r-1})=\z_{n_1,\ldots,n_r}$, the shuffle product
$\z(w_1)\z(w_2)=\z(w_1\shuffle w_2)$ and $\z(e_0)=0=\z(e_1)$.

The Deligne canonical associator $W$ is related to the Drinfeld associator $Z$ 
through the following equation \SVMZV
\eqn\Delignea{
W\circ {}^\sigma Z=Z\ ,}
with the Ihara action $\circ$ and the anti--linear map $\sigma: \IC\vev{e_0,e_1}\ra \IC\vev{e_0,e_1}$ with $\sigma(e_i)\mapsto -e_i$.
The equation \Delignea\ can be solved  recursively in length of words as \SVMZV:
\eqn\recursive{
W(e_0,e_1)={}^\sigma Z(e_0,We_1W^{-1})^{-1}\ Z(e_0,e_1)\ .}
The unique solution to \fixed\ is $e_1'=We_1W^{-1}$. As a consequence, \eqqs \genLc\ and 
\recursive\ allow to express the Deligne associator $W$ as \SVMZV:
\eqn\DeligneA{
\Lc(1)=W(e_0,e_1)=Z(-e_0,-e_1')^{-1}\ Z(e_0,e_1)\ .}

In analogy to the motivic version of the Drinfeld associator \Drinfeld
\eqn\MotDrinfeld{
Z^m(e_0,e_1)=\sum_{w\in\{e_0,e_1\}^\times}\z^m(w)\ w}
in {\it Ref.} \SVMZV\ Brown has given  the motivic single--valued associator as a generating series
\eqn\motDeligne{
W^m(e_0,e_1)=\sum_{w\in\{e_0,e_1\}^\times}\z_{\rm sv}^m(w)\ w\ ,}
whose period map $\per$ gives the Deligne associator $W(e_0,e_1)$.
Note, that the motivic SVMZVs $\z_{\rm sv}^m(w)$ satisfy the same double shuffle and associator relations than the motivic MZVs $\z^m(w)$.
Hence, in a first step we can work out the sum \motDeligne\ in the same way as \MotDrinfeld\ by applying various shuffle and associator relations,   
in the second step we replace the symbols $\z^m(w)$ by $\z_{\rm sv}^m(w)$. Finally, in the last step the latter are replaced thanks to  relations such as (the motivic versions of) \Example\ and 
\Examplee.  As a result we obtain:
\eqn\MotDeligne{\eqalign{
W^m(e_0,e_1)&=1+2\ \z_3^m\ ([e_0,[e_0,e_1]]-[e_1,[e_0,e_1])+2\ \z_5^m\ 
\Big([e_0,[e_0,[e_0,[e_0,e_1]]]]\cr
&-\h\ [e_0,[e_0,[e_1,[e_0,e_1]]]]-\fc{3}{2}\ [e_1,[e_0,[e_0,[e_0,e_1]]]]
+(e_0\leftrightarrow e_1)\Big)+\ldots\ .}}

The associator $Z$ is group--like. Therefore, its logarithm $\ln Z$ can be expressed as a Lie series in the elements $e_0$ and $e_1$. Due to Drinfeld we have \Drinii
\eqn\logDrinfeld{
\ln Z(e_0,e_1)=\sum_{k,l\geq 1} z_{kl}\ u_{kl}\ \ \mod\ g''\ ,}
with ($\ad_xy=[x,y]$)
\eqn\Coeff{
u_{kl}=(-1)^k\ \ad_{e_1}^{k-1}\ \ad^{l-1}_{e_0}[e_0,e_1]\ ,}
and $g''=[[g,g],[g,g]]$  the second commutant of the Lie algebra $g$. The 
coefficients $z_{kl}$ are extracted from the generating function
\eqn\coeffDrinfeld{\eqalign{
\fc{\Gamma(1-u)\ \Gamma(1-v)}{\Gamma(1-u-v)}&=1+\sum_{k,l\geq 1} z_{kl}\ u^k\ v^l\cr
&=1-\z_2\ u\;v-\z_3\ u\;v(u+v)-\z_4\ u\;v(u^2+\fc{1}{4}uv+v^2)+\ldots ,}}
which in turn gives rise to
\eqn\LogDrinfeld{
\ln Z(e_0,e_1)=-(uv)^{-1}\ \lf(\fc{\Gamma(1-u)\ \Gamma(1-v)}{\Gamma(1-u-v)}-1\ri) [e_0,e_1]
\ \ \mod\ g''\ ,}
with:
\eqn\rule{
u=-\ad_{e_1}\ ,\ \ \ v=\ad_{e_0}\ .}
Furthermore, we have
\eqn\myDrinfeld{
Z(e_0,e_1)=1-(uv)^{-1}\ \lf(\fc{\Gamma(1-u)\ \Gamma(1-v)}{\Gamma(1-u-v)}-1\ri) [e_0,e_1]
\ \ \mod\ (g')^2\ ,}
which reproduces \Drinfeld\ up to squares of commutators $(g')^2=[g,g]^2$.

In {\it Ref.} \Drummond\ by making use of the Ihara bracket \Ihara
\eqn\Iharabr{
\{x,y\}=[x,y]+D_xy-D_yx\ ,\ \ \ x,y\in g\ ,}
with the derivations 
\eqn\derivations{
D_y e_0=0\ \ \ ,\ \ \ \ D_y e_1=[e_1,y]\ ,}
the Drinfeld associator \Drinfeld\ has been written in a form, which very much resembles the structure of the period 
matrix $F$, given in \period. In particular, the terms $Q_n$ containing the 
MZVs of depth greater than one are accompanied by Ihara brackets.
In a limit, where the latter vanishes, a correspondence can be established 
between the four--point open superstring amplitude $\Ac$
\eqn\fourg{
\Ac(1,2,3,4)=\fc{\Gamma(1+s)\ \Gamma(1+u)}{\Gamma(1+s+u)}\ A\ ,}
with the two kinematic invariants $s= \alpha'(k_1+k_2)^2$ and $u= \alpha'(k_1+k_4)^2$
and the associator.
If we work modulo $g''$ then a commutative realization of the Ihara bracket is established
and \fourg\ can be related to \LogDrinfeld\ \Drummond.

A natural question is, whether the four--point closed superstring amplitude \Graviton\
can be related to the Deligne associator $W(e_0,e_1)$.
The four--graviton amplitude \Graviton
\eqn\fourG{
\Mc(1,2,3,4)=G\ |A|^2\ ,}
can be obtained from \Gmatrix\ 
\eqn\four{
G=S_0\ \exp\lf\{-2\ \sum_{n\geq 1}\zeta_{2n+1}\ M_{2n+1}\ri\}\ ,}
with
\eqn\fourm{
M_{2n+1}=-\fc{1}{2n+1}\ \lf[\ s^{2n+1}+u^{2n+1}-(s+u)^{2n+1}\ \ri]\ ,}
and the normalization:
\eqn\Intersection{
S_0=-\pi\  \fc{su}{s+u}\ .}
Hence, we have:
\eqn\Four{
\Mc(1,2,3,4)=|S_0|\ 
\fc{\Gamma(s)\ \Gamma(u)\ \Gamma(-s-u)}{\Gamma(-s)\ \Gamma(-u)\ \Gamma(s+u)}\ |A|^2\ .}
In the case under consideration the expansion \gravlowmom\ of \four\ becomes rather
simple, all $Q_{\rm sv}(n)$ disappear and the commutator brackets become trivial.

By borrowing the arguments given below \eqq \motDeligne\ the (motivic) Deligne associator 
\motDeligne\ can also be cast in a form, where MZVs of depth greater than one are accompanied 
by Ihara brackets. Note, that this step essentially amounts to replacing $\z^m$ by $\z_{\rm sv}^m$
in the (motivic) Drinfeld associator written in terms of the Ihara bracket 
along the lines of \Drummond. After these steps we find
\eqnn\applyDrummond{
$$\eqalignno{
W^m(e_0,e_1) &= 1 + \SVM(3)\ w_3 + \SVM(5)\ w_5  + \h\ \SVM(3)^2\ (w_3 \circ w_3)+ 
\SVM(7)\ w_7 \cr
& + \fc{1}{5}\ \SVM(3,5)\ \{ w_5,w_3\} +\SVM(3)\ \SVM(5)\ (w_5 \circ w_3)+ \SVM(9)\ w_9\cr
& + \fc{1}{3!}\ \SVM(3)^3\ ((w_3 \circ w_3) \circ w_3)+ 
\lf(\fc{3}{14}\  \SVM(5)^2 + \fc{1}{14}\ \SVM(3,7)\ri) \{w_7,w_3\}\cr
&  + \SVM(3)\ \SVM(7)\  (w_7 \circ w_3) +\h\ \SVM(5)^2\ (w_5 \circ w_5) 
+\ldots\ ,&\applyDrummond}$$}
with the abbreviations \Drummond:
\eqnn\WS{
$$\eqalignno{
w_3 &= [e_0 ,[e_0,e_1]]-[e_1,[e_0,e_1]]\ ,\cr
w_5 &= [e_0,[e_0,[e_0,[e_0,e_1]]]] - \h\ [e_0,[e_0,[e_1,[e_0,e_1]]]] - 
\fc{3}{2}\ [e_1,[e_0,[e_0,[e_0,e_1]]]] \cr
& + (e_0 \leftrightarrow e_1)\ .&\WS}$$}
In \applyDrummond\ we have introduced  the right action of $g$ as:
\eqn\rightaction{
x \circ y=xy-D_y x\ \ \ ,\ \ \ y\in g,\ x\in U(g)\ .}
Obviously, in this form \applyDrummond\ the Deligne associator is very similar to $\sv(F^m)$, given in \applyF, \ie $W^m\simeq \sv(F^m)$.
Comparing \applyDrummond\ and \applyF\ amounts to replace the words $w_{2r+1}$ by the matrices 
$M_{2r+1}$ and the action $\circ$ by the matrix product. The Ihara bracket 
$\{w_{2r_1+1},w_{2r_2+1}\}=w_{2r_1+1}\circ w_{2r_2+1}-w_{2r_2+1}\circ w_{2r_2+1}$ is replaced by matrix commutator $[M_{2r_1+1},M_{2r_2+1}]$.
Hence, to find a correspondence between \Four\ and the Deligne associator, we also 
need to work in a commutative realization of the Ihara bracket, \ie modulo~$g''$.

In this limit, based on the closed superstring amplitude \Four\ we conjecture the following expression for the Deligne associator $W(e_0,e_1)$
\eqn\logDeligne{
\ln W(e_0,e_1)=\sum_{k,l\geq 1} w_{kl}\ u_{kl}\ \ \mod\ g'',}
with the coefficients $w_{kl}$ extracted from the generating function:
\eqn\coeffDeligne{\eqalign{
-\fc{\Gamma(-u)\ \Gamma(-v)\ \Gamma(u+v)}{\Gamma(u)\ \Gamma(v)\ \Gamma(-u-v)}&=1+\sum_{k,l\geq 1} w_{kl}\ u^k\ v^l\cr
&=1-2\ \z_3\ u\;v(u+v)-2\ \z_5\ u\;v(u+v)(u^2+uv+v^2)\cr
&+2\ \z_3^2\; u^2\ v^2(u+v)^2+\ldots\ .}}
With \coeffDeligne\ the sum \logDeligne\ gives:
\eqn\logdeligne{
\ln W(e_0,e_1)=2\ \z_3\ ([e_0,[e_0,e_1]]-[e_1,[e_0,e_1])+\ldots \ .}
{\it Eqs.} \logDeligne\ and \coeffDeligne\ can be combined into:
\eqn\LogDeligne{
\ln W(e_0,e_1)=(uv)^{-1}\ \lf(\fc{\Gamma(-u)\ \Gamma(-v)\ \Gamma(u+v)}{\Gamma(u)\ \Gamma(v)\ \Gamma(-u-v)}+1\ri) [e_0,e_1]\ \ \mod\ g''\ ,}
which gives rise to:
\eqn\myDeligne{
W(e_0,e_1)=1+(uv)^{-1}\ \lf(\fc{\Gamma(-u)\ \Gamma(-v)\ \Gamma(u+v)}{\Gamma(u)\ \Gamma(v)\ \Gamma(-u-v)}+1\ri) [e_0,e_1]\ \ \mod\ (g')^2\ .}
Expanding \myDeligne\ w.r.t. $u$ and $v$ reproduces the Deligne associator $W(e_0,e_1)$ 
modulo\foot{Note, that expanding \myDeligne\ yields the $\z_5$-- and $\z_3^2$--terms  
displayed, which modulo squares of commutators $(g')^2$ agree with the exact terms:
$2\z_5\Big([e_0,[e_0,[e_0,[e_0,e_1]]]]-\h [e_0,[e_0,[e_1,[e_0,e_1]]]]-\fc{3}{2} [e_1,[e_0,[e_0,[e_0,e_1]]]]
+(e_0\leftrightarrow e_1)\Big)$ and $2\z_3^2\Big(w_3^2-[e_0,[e_0,[e_1,w_3]]]+[e_1,[e_0,[e_1,w_3]]]+
[[e_1,w_3],[e_0,e_1]]\Big)$, respectively. Here, $w_3$ is defined in \WS.}
squares of commutators $(g')^2$:
\eqn\Deligne{\eqalign{
W(e_0,e_1)&=1+2\ \z_3\ ([e_0,[e_0,e_1]]-[e_1,[e_0,e_1])+2\ \z_5\ \Big([e_0,[e_0,[e_0,[e_0,e_1]]]]
\cr
&-2\ [e_1,[e_0,[e_0,[e_0,e_1]]]]+2\ [e_1,[e_1,[e_0,[e_0,e_1]]]]-[e_1,[e_1,[e_1,[e_0,e_1]]]]\Big)\cr
&+2\ \z_3^2\ \Big([e_1,[e_1,[e_1,[e_0,[e_0,e_1]]]]]+[e_1,[e_0,[e_0,[e_0,[e_0,e_1]]]]]\cr
&-2\ [e_1,[e_1,[e_0,[e_0,[e_0,e_1]]]]]\Big)+\ldots\ .}}
Note, that the form \Deligne\ agrees with the motivic version \MotDeligne\ subject to the period
map \per.
It can be verified, that the expressions \LogDeligne\ and \LogDrinfeld\ indeed fulfill \DeligneA, \ie
\eqn\fulfill{
\ln W(e_0,e_1)=-\ln Z(-e_0,-e_1')+\ln Z(e_0,e_1)\ \ \mod\ g'',}
modulo double commutators and $e_1'=We_1W^{-1}$. This relation \fulfill\ can be checked 
order by order in a basis of MZVs.

To conclude, the motivic Deligne associator \MotDeligne\ written explicitly as \applyDrummond\ assumes the same formal expansion \applyF\ as the relevant piece of the closed superstring 
amplitude, \ie $W^m\simeq \sv(F^m)$.
Furthermore,  the Deligne associator \myDeligne\ (modulo squares of commutators $(g')^2$) assumes a similar form as the four--graviton amplitude \Four\ just as the Drinfeld associator resembles the four--gluon amplitude \Drummond.
The relation \Delignea\ should be  interpreted as Kawai--Lewellen--Tye (KLT) relation for the associators.
It should be straightforward to use the Deligne associator \DeligneA\ as a tool for setting up
recursion relations for general $N$--graviton amplitudes in lines of \Broedel.

\newsec{Concluding remarks}

In this work  we have revisited  the $\ap$--expansion of the  closed
superstring amplitude (graviton amplitude) at tree--level with particular emphasis on its  underlying algebraic structure and transcendentality properties.

We have found a striking similarity \Wefind\ between the open and closed superstring amplitudes communicated by the homomorphism  \mapSV.
After mapping the motivic open and closed superstring amplitudes onto a non--commutative Hopf--algebra this analogy is given by \RESULTT\ and established by the map  \mapsv.
In other words, through the relations \Wefind\ or \RESULTT\ the closed superstring amplitude is obtained from the open superstring amplitude by some truncation realized by the maps
\mapSV\ or \mapsv, respectively. 
In the writings \canbewritten\ and \GRAVITON\ (or \motO\ and \motC) the $\ap$--expansions of the  tree--level open and closed superstring amplitude take an uniform form suggesting an even deeper connection between gauge and gravity amplitudes than what is implied by KLT relations \KawaiXQ.
Anyhow, an apparent similarity  between perturbative gauge-- and gravity--theories
is established in field--theory through the double copy construction \BCJ\ 
and in string theory through the Mellin correspondence furnishing a superstring/supergravity 
 resemblance \StiebergerHZA.
Furthermore, recently interesting uniform descriptions of gauge-- and gravity amplitudes
in field--theory have been presented in \CachazoIEA.

The relation \Wefind\ relates two very different superstring amplitudes by the map \mapSV.
It would be interesting to understand the role of the map $\sv$ at the level of the perturbation
theory of open and closed strings or from the nature of the underlying string world--sheets. 
Note, that a large class of Feynman integrals in four space--time dimensions 
lives in the subspace of SVMZVs or SVMPs, \cf {\it Refs.} \doubref\Duhr\LeurentMR. 
As pointed out by Brown in \SVMZV, this fact opens the interesting possibility
to replace general amplitudes  with their single--valued versions (defined by the map $\sv$),
which should lead to considerable simplifications.
In string theory this simplification occurs by replacing gluon amplitudes \canbewritten\ 
(or \motO) with their single--valued versions describing
graviton amplitudes \GRAVITON\ (or \motC), which seem to be considerably simpler.

We have identified the Deligne associator \MotDeligne\ to carry the relevant information on the
closed superstring amplitude.  Modulo squares of commutators for the Deligne associator 
we have presented a closed form \myDeligne\ in terms of Gamma functions
 and have argued, that it is related to the four--point closed superstring amplitude very much in the sense as the Drinfeld associator can be related to
the four--point open superstring amplitude \Drummond.
Hence, the relation \recursive, which defines the Deligne associator as a product of two Drinfeld associators (at different arguments), might be interpreted as sort of KLT relation for associators.
It would certainly be interesting to understand better such an interpretation.

Through the $\Gamma_\sigma$--decomposition $B_\sigma(s,u)=\fc{\Gamma(s)\Gamma(u)}{\Gamma(s+u)},\ \sigma\in GT$ the Drinfeld associator \LogDrinfeld\  is related to the structure of the Lie algebra of the Grothendieck--Teichm\"uller group 
$GT$ \Iharai. The latter
plays a role in revealing the underlying Lie algebra structure of the open superstring amplitude. Hence, the Deligne associator \LogDeligne\ should be related to the underlying algebra of 
the $\ap$--expansion of closed superstring amplitude, \cf also {\it Refs.} 
\seealso\ for related research.

Finally, the structure underlying the motivic open and closed superstring amplitudes
in terms of a Hopf algebra is not only a tool to conveniently express these amplitudes
but rather seems to be an intrinsic feature, which might allow to compute the latter
by first principles. Eventually, some or all aspects of string perturbation
theory might be reduced to algebraic methods based on arithmetic algebraic geometry.

\vskip2.5cm
\goodbreak
\leftline{\noindent{\bf Acknowledgments}}

\noindent
I wish to thank  Hidekazu Furusho for useful comments.

\listrefs

\end

%% file: myharvmac
%%%%%%%%%%%%%%%%%%  tex macros for preprints, cm version %%%%%%%%%%%%%%
%         (P. Ginsparg <ginsparg@lanl.gov>, last updated 7/94)
%                if confused, type `b' in response to query 
%           hypertex extensions (still provisional), 7/26/94
%
%---------------------------------------------------------------------%
%\input hyperbasics %comment out this line to restore non-hyper functionality
%
%% site dependent options:
%% \unredoffs and \redoffs define horizontal and vertical offsets
%% respectively for unreduced and reduced modes. \speclscape defines
%% the \special{} call that sets printer to landscape (sideways) mode.
%% from standard set below, leave uncommented as appropriate or redefine
%
%%% next 400dpi
\def\unredoffs{} \def\redoffs{\voffset=-.31truein\hoffset=-.48truein}
\def\speclscape{}
%\def\speclscape{\special{papersize=11in,8.5in}}
%
%%% apple lw
%\def\unredoffs{} \def\redoffs{\voffset=-.31truein\hoffset=-.59truein}
%\def\speclscape{\special{ps: landscape}}
%
%%% qms lasergrafix:
%\def\unredoffs{} \def\redoffs{\voffset=-.4truein\hoffset=.125truein}
%\def\speclscape{\special{qms: landscape}}
%
%%% saclay A4 paper:
%\def\unredoffs{\hoffset-.14truein\voffset-.2truein}
%\def\redoffs{\voffset=-.45truein\hoffset=-.21truein}
%\def\speclscape{\special{landscape}}
%
%---------------------------------------------------------------------%
%
\newbox\leftpage \newdimen\fullhsize \newdimen\hstitle \newdimen\hsbody
\tolerance=1000\hfuzz=2pt
\catcode`\@=11 % This allows us to modify PLAIN macros.
\ifx\hyperdef\UNd@FiNeD\def\hyperdef#1#2#3#4{#4}\def\hyperref#1#2#3#4{#4}\fi
\def\bigans{b }
\def\answ{b }
%\message{ big or little (b/l)? }\read-1 to\answ
%
\ifx\answ\bigans\message{(This will come out unreduced.}
\magnification=1200\unredoffs\baselineskip=16pt plus 2pt minus 1pt
\hsbody=\hsize \hstitle=\hsize %take default values for unreduced format
\else\message{(This will be reduced.} \let\l@r=L
\magnification=1000\baselineskip=16pt plus 2pt minus 1pt \vsize=7truein
\redoffs \hstitle=8truein\hsbody=4.75truein\fullhsize=10truein\hsize=\hsbody
\output={\ifnum\pageno=0 %%% This is the HUTP version
  \shipout\vbox{\speclscape{\hsize\fullhsize\makeheadline}
    \hbox to \fullhsize{\hfill\pagebody\hfill}}\advancepageno
  \else
  \almostshipout{\leftline{\vbox{\pagebody\makefootline}}}\advancepageno
  \fi}
\def\almostshipout#1{\if L\l@r \count1=1 \message{[\the\count0.\the\count1]}
      \global\setbox\leftpage=#1 \global\let\l@r=R
 \else \count1=2
  \shipout\vbox{\speclscape{\hsize\fullhsize\makeheadline}
      \hbox to\fullhsize{\box\leftpage\hfil#1}}  \global\let\l@r=L\fi}
\fi
%---------------------------------------------------------------------
%
\newcount\yearltd\yearltd=\year\advance\yearltd by -2000

\def\Title#1#2{\nopagenumbers\abstractfont\hsize=\hstitle\rightline{#1}%
\vskip 1in\centerline{\titlefont #2}\abstractfont\vskip .5in\pageno=0}
\def\Date#1{\vfill\leftline{#1}\tenpoint\supereject\global\hsize=\hsbody%
\footline={\hss\tenrm\hyperdef\hypernoname{page}\folio\folio\hss}}%
% (restores pagenumbers)
%
%       use following instead of \Date on the preliminary draft,
%       puts date/time on each page in big mode, writes labels in margins

\def\draftmode{\message{ DRAFTMODE }\def\draftdate{{\rm preliminary draft:
\number\month/\number\day/{0}\number\yearltd\ \ \hourmin}}%
\headline={\hfil\draftdate}\writelabels\baselineskip=20pt plus 2pt minus 2pt
 {\count255=\time\divide\count255 by 60 \xdef\hourmin{\number\count255}
  \multiply\count255 by-60\advance\count255 by\time
  \xdef\hourmin{\hourmin:\ifnum\count255<10 0\fi\the\count255}}}
%       use \nolabels to get rid of eqn, ref, and fig labels in draft mode
\def\nolabels{\def\wrlabeL##1{}\def\eqlabeL##1{}\def\reflabeL##1{}}
\def\writelabels{\def\wrlabeL##1{\leavevmode\vadjust{\rlap{\smash%
{\line{{\escapechar=` \hfill\rlap{\sevenrm\hskip.03in\string##1}}}}}}}%
\def\eqlabeL##1{{\escapechar-1\rlap{\sevenrm\hskip.05in\string##1}}}%
\def\reflabeL##1{\noexpand\llap{\noexpand\sevenrm\string\string\string##1}}}
\nolabels
%
% tagged sec numbers
\global\newcount\secno \global\secno=0
\global\newcount\meqno \global\meqno=1
\def\s@csym{}
\def\newsec#1{\global\advance\secno by1%
{\toks0{#1}\message{(\the\secno. \the\toks0)}}%
%\ifx\answ\bigans \vfill\eject \else \bigbreak\bigskip \fi  %if desired
\global\subsecno=0\eqnres@t\let\s@csym\secsym\xdef\secn@m{\the\secno}\noindent
{\bf\hyperdef\hypernoname{section}{\the\secno}{\the\secno.} #1}%
\writetoca{{\string\hyperref{}{section}{\the\secno}{\it\the\secno.}} {{\it #1} }}%
\par\nobreak\medskip\nobreak}
\def\eqnres@t{\xdef\secsym{\the\secno.}\global\meqno=1\bigbreak\bigskip}
\def\sequentialequations{\def\eqnres@t{\bigbreak}}\xdef\secsym{}
\global\newcount\subsecno \global\subsecno=0
\def\subsec#1{\global\advance\subsecno by1%
{\toks0{#1}\message{(\s@csym\the\subsecno. \the\toks0)}}%
\ifnum\lastpenalty>9000\else\bigbreak\fi       \global\subsubsecno=0
\noindent{\it\hyperdef\hypernoname{subsection}{\secn@m.\the\subsecno}%
{\secn@m.\the\subsecno.} #1}\writetoca{\string\quad
{\string\hyperref{}{subsection}{\secn@m.\the\subsecno}{\secn@m.\the\subsecno.}}
{#1}}\par\nobreak\medskip\nobreak}
\def\appendix#1#2{\global\meqno=1\global\subsecno=0\xdef\secsym{\hbox{#1.}}%
\bigbreak\bigskip\noindent{\bf Appendix \hyperdef\hypernoname{appendix}{#1}%
{#1.} #2}{\toks0{(#1. #2)}\message{\the\toks0}}%
\xdef\s@csym{#1.}\xdef\secn@m{#1}%
\writetoca{\string\hyperref{}{appendix}{#1}{{\it Appendix} {\it #1.}} {\it #2}}%
\par\nobreak\medskip\nobreak}
%
%       \eqn\label{a+b=c}	gives displayed equation, numbered
%				consecutively within sections.
%     \eqnn and \eqna define labels in advance (of eqalign?)
%
\def\checkm@de#1#2{\ifmmode{\def\f@rst##1{##1}\hyperdef\hypernoname{equation}%
{#1}{#2}}\else\hyperref{}{equation}{#1}{#2}\fi}
\def\eqnn#1{\DefWarn#1\xdef #1{(\noexpand\relax\noexpand\checkm@de%
{\s@csym\the\meqno}{\secsym\the\meqno})}%
\wrlabeL#1\writedef{#1\leftbracket#1}\global\advance\meqno by1}
\def\f@rst#1{\c@t#1a\em@ark}\def\c@t#1#2\em@ark{#1}
\def\eqna#1{\DefWarn#1\wrlabeL{#1$\{\}$}%
\xdef #1##1{(\noexpand\relax\noexpand\checkm@de%
{\s@csym\the\meqno\noexpand\f@rst{##1}}{\hbox{$\secsym\the\meqno##1$}})}
\writedef{#1\numbersign1\leftbracket#1{\numbersign1}}\global\advance\meqno by1}
\def\eqn#1#2{\DefWarn#1%
\xdef #1{(\noexpand\hyperref{}{equation}{\s@csym\the\meqno}%
{\secsym\the\meqno})}$$#2\eqno(\hyperdef\hypernoname{equation}%
{\s@csym\the\meqno}{\secsym\the\meqno})\eqlabeL#1$$%
\writedef{#1\leftbracket#1}\global\advance\meqno by1}
\def\xeqn{\expandafter\xe@n}\def\xe@n(#1){#1}
\def\xeqna#1{\expandafter\xe@n#1}
\def\eqns#1{(\e@ns #1{\hbox{}})}
\def\e@ns#1{\ifx\UNd@FiNeD#1\message{eqnlabel \string#1 is undefined.}%
\xdef#1{(?.?)}\fi{\let\hyperref=\relax\xdef\next{#1}}%
\ifx\next\em@rk\def\next{}\else%
\ifx\next#1\xeqn#1\else\def\n@xt{#1}\ifx\n@xt\next#1\else\xeqna#1\fi
\fi\let\next=\e@ns\fi\next}

\def\DefWarn#1{\ifx\UNd@FiNeD#1\else
\immediate\write16{*** WARNING: the label \string#1 is already defined ***}\fi}
%
%			 footnotes
\newskip\footskip\footskip14pt plus 1pt minus 1pt %sets footnote baselineskip
\def\footnotefont{\ninepoint}\def\f@t#1{\footnotefont #1\@foot}
\def\f@@t{\baselineskip\footskip\bgroup\footnotefont\aftergroup\@foot\let\next}
\setbox\strutbox=\hbox{\vrule height9.5pt depth4.5pt width0pt}
\global\newcount\ftno \global\ftno=0
\def\foot{\global\advance\ftno by1\def\foot@rg{\hyperref{}{footnote}%
{\the\ftno}{\the\ftno}\xdef\foot@rg{\noexpand\hyperdef\noexpand\hypernoname%
{footnote}{\the\ftno}{\the\ftno}}}\footnote{$^{\foot@rg}$}}
%
%say \footend to put footnotes at end
%will cause problems if \ref used inside \foot, instead use \nref before
\newwrite\ftfile
\def\footend{\def\foot{\global\advance\ftno by1\chardef\wfile=\ftfile
%%$^{\the\ftno}$\ifnum\ftno=1\immediate\openout\ftfile=\jobname.fts\fi%
\hyperref{}{footnote}{\the\ftno}{$^{\the\ftno}$}%
\ifnum\ftno=1\immediate\openout\ftfile=\jobname.fts\fi%
\immediate\write\ftfile{\noexpand\smallskip%
%%\noexpand\item{f\the\ftno:\ }\pctsign}\findarg}%
\noexpand\item{\noexpand\hyperdef\noexpand\hypernoname{footnote}
{\the\ftno}{f\the\ftno}:\ }\pctsign}\findarg}%
\def\footatend{\vfill\eject\immediate\closeout\ftfile{\parindent=20pt
\centerline{\bf Footnotes}\nobreak\bigskip\input \jobname.fts }}}
\def\footatend{}
%
%     \ref\label{text}
% generates a number, assigns it to \label, generates an entry.
% To list the refs on a separate page,  \listrefs
%
\global\newcount\refno \global\refno=1
\newwrite\rfile
\def\ref{[\hyperref{}{reference}{\the\refno}{\the\refno}]\nref}
\def\nref#1{\DefWarn#1%
\xdef#1{[\noexpand\hyperref{}{reference}{\the\refno}{\the\refno}]}%
\writedef{#1\leftbracket#1}%
\ifnum\refno=1\immediate\openout\rfile=\jobname.refs\fi
\chardef\wfile=\rfile\immediate\write\rfile{\noexpand\item{[\noexpand\hyperdef%
\noexpand\hypernoname{reference}{\the\refno}{\the\refno}]\ }%
\reflabeL{#1\hskip.31in}\pctsign}\global\advance\refno by1\findarg}
%	horrible hack to sidestep tex \write limitation
\def\findarg#1#{\begingroup\obeylines\newlinechar=`\^^M\pass@rg}
{\obeylines\gdef\pass@rg#1{\writ@line\relax #1^^M\hbox{}^^M}%
\gdef\writ@line#1^^M{\expandafter\toks0\expandafter{\striprel@x #1}%
\edef\next{\the\toks0}\ifx\next\em@rk\let\next=\endgroup\else\ifx\next\empty%
\else\immediate\write\wfile{\the\toks0}\fi\let\next=\writ@line\fi\next\relax}}
\def\striprel@x#1{} \def\em@rk{\hbox{}}
\def\lref{\begingroup\obeylines\lr@f}
\def\lr@f#1#2{\DefWarn#1\gdef#1{\let#1=\UNd@FiNeD\ref#1{#2}}\endgroup\unskip}

\def\addref#1{\immediate\write\rfile{\noexpand\item{}#1}} %now unnecessary
\def\listrefs{\footatend\vfill\supereject\immediate\closeout\rfile\writestoppt
\baselineskip=\footskip\centerline{{\bf References}}\bigskip{\parindent=20pt%
\frenchspacing\escapechar=` \input \jobname.refs\vfill\eject}\nonfrenchspacing}
\def\startrefs#1{\immediate\openout\rfile=\jobname.refs\refno=#1}
\def\xref{\expandafter\xr@f}\def\xr@f[#1]{#1}
\def\refs#1{\count255=1[\r@fs #1{\hbox{}}]}
\def\r@fs#1{\ifx\UNd@FiNeD#1\message{reflabel \string#1 is undefined.}%
\nref#1{need to supply reference \string#1.}\fi%
\vphantom{\hphantom{#1}}{\let\hyperref=\relax\xdef\next{#1}}%
\ifx\next\em@rk\def\next{}%
\else\ifx\next#1\ifodd\count255\relax\xref#1\count255=0\fi%
\else#1\count255=1\fi\let\next=\r@fs\fi\next}
%

%
% this is ugly, but moore insists
\newwrite\ffile\global\newcount\figno \global\figno=1
\def\fig{fig.~\hyperref{}{figure}{\the\figno}{\the\figno}\nfig}
\def\nfig#1{\DefWarn#1%
\xdef#1{fig.~\noexpand\hyperref{}{figure}{\the\figno}{\the\figno}}%
\writedef{#1\leftbracket fig.\noexpand~\xfig#1}%
\ifnum\figno=1\immediate\openout\ffile=\jobname.figs\fi\chardef\wfile=\ffile%
{\let\hyperref=\relax
\immediate\write\ffile{\noexpand\medskip\noexpand\item{Fig.\ %
\noexpand\hyperdef\noexpand\hypernoname{figure}{\the\figno}{\the\figno}. }
\reflabeL{#1\hskip.55in}\pctsign}}\global\advance\figno by1\findarg}
\def\listfigs{\vfill\eject\immediate\closeout\ffile{\parindent40pt
\baselineskip14pt\centerline{{\bf Figure Captions}}\nobreak\medskip
\escapechar=` \input \jobname.figs\vfill\eject}}
\def\xfig{\expandafter\xf@g}\def\xf@g fig.\penalty\@M\ {}
\def\figs#1{figs.~\f@gs #1{\hbox{}}}
\def\f@gs#1{{\let\hyperref=\relax\xdef\next{#1}}\ifx\next\em@rk\def\next{}\else
\ifx\next#1\xfig #1\else#1\fi\let\next=\f@gs\fi\next}
\def\figin{\epsfcheck\figin}\def\figins{\epsfcheck\figins}
\def\epsfcheck{\ifx\epsfbox\UNd@FiNeD
\message{(NO epsf.tex, FIGURES WILL BE IGNORED)}
\gdef\figin##1{\vskip2in}\gdef\figins##1{\hskip.5in}% blank space instead
\else\message{(FIGURES WILL BE INCLUDED)}%
\gdef\figin##1{##1}\gdef\figins##1{##1}\fi}
\def\DefWarn#1{}
\def\figinsert{\goodbreak\midinsert}
\def\ifig#1#2#3{\DefWarn#1\xdef#1{Fig.~\noexpand\hyperref{}{figure}%
{\the\figno}{\the\figno}}\writedef{#1\leftbracket fig.\noexpand~\xfig#1}%
\figinsert\figin{\centerline{#3}}\medskip\centerline{\vbox{\baselineskip12pt
\advance\hsize by -1truein\noindent\wrlabeL{#1=#1}\footnotefont%
{\bf Fig.~\hyperdef\hypernoname{figure}{\the\figno}{\the\figno}:} #2}}
\bigskip\endinsert\global\advance\figno by1}
\newwrite\lfile
{\escapechar-1\xdef\pctsign{\string\%}\xdef\leftbracket{\string\{}
\xdef\rightbracket{\string\}}\xdef\numbersign{\string\#}}
\def\writedefs{\immediate\openout\lfile=\jobname.defs \def\writedef##1{%
{\let\hyperref=\relax\let\hyperdef=\relax\let\hypernoname=\relax
 \immediate\write\lfile{\string\def\string##1\rightbracket}}}}%
\def\writestop{\def\writestoppt{\immediate\write\lfile{\string\pageno
 \the\pageno\string\startrefs\leftbracket\the\refno\rightbracket
 \string\def\string\secsym\leftbracket\secsym\rightbracket
 \string\secno\the\secno\string\meqno\the\meqno}\immediate\closeout\lfile}}
\def\writestoppt{}\def\writedef#1{}
\def\seclab#1{\DefWarn#1%
\xdef #1{\noexpand\hyperref{}{section}{\the\secno}{\the\secno}}%
\writedef{#1\leftbracket#1}\wrlabeL{#1=#1}}
\def\subseclab#1{\DefWarn#1%
\xdef #1{\noexpand\hyperref{}{subsection}{\secn@m.\the\subsecno}%
{\secn@m.\the\subsecno}}\writedef{#1\leftbracket#1}\wrlabeL{#1=#1}}
\def\applab#1{\DefWarn#1%
\xdef #1{\noexpand\hyperref{}{appendix}{\secn@m}{\secn@m}}%
\writedef{#1\leftbracket#1}\wrlabeL{#1=#1}}
\newwrite\tfile \def\writetoca#1{}
\def\leaderfill{\leaders\hbox to 1em{\hss.\hss}\hfill}
%	use this to write file with table of contents
\def\writetoc{\immediate\openout\tfile=\jobname.toc
   \def\writetoca##1{{\edef\next{\write\tfile{\noindent ##1
   \string\leaderfill {\string\hyperref{}{page}{\noexpand\number\pageno}%
                       {\noexpand\number\pageno}} \par}}\next}}}
%       and this lists table of contents on second pass
\newread\ch@ckfile
\def\listtoc{\immediate\closeout\tfile\immediate\openin\ch@ckfile=\jobname.toc
\ifeof\ch@ckfile\message{no file \jobname.toc, no table of contents this pass}%
\else\closein\ch@ckfile\centerline{\bf Contents}\nobreak\medskip%
{\baselineskip=18.5pt  \footnotefont
\parskip=2pt\catcode`\@=12\input\jobname.toc
\catcode`\@=12\bigbreak\bigskip}\fi}
\catcode`\@=12 % at signs are no longer letters
%
%	Unpleasantness in calling in abstract and title fonts
\edef\tfontsize{\ifx\answ\bigans scaled\magstep3\else scaled\magstep4\fi}
\font\titlerm=cmr10 \tfontsize \font\titlerms=cmr7 \tfontsize
\font\titlermss=cmr5 \tfontsize \font\titlei=cmmi10 \tfontsize
\font\titleis=cmmi7 \tfontsize \font\titleiss=cmmi5 \tfontsize
\font\titlesy=cmsy10 \tfontsize \font\titlesys=cmsy7 \tfontsize
\font\titlesyss=cmsy5 \tfontsize \font\titleit=cmti10 \tfontsize
\skewchar\titlei='177 \skewchar\titleis='177 \skewchar\titleiss='177
\skewchar\titlesy='60 \skewchar\titlesys='60 \skewchar\titlesyss='60
\def\titlefont{\def\rm{\fam0\titlerm}% switch to title font
\textfont0=\titlerm \scriptfont0=\titlerms \scriptscriptfont0=\titlermss
\textfont1=\titlei \scriptfont1=\titleis \scriptscriptfont1=\titleiss
\textfont2=\titlesy \scriptfont2=\titlesys \scriptscriptfont2=\titlesyss
\textfont\itfam=\titleit \def\it{\fam\itfam\titleit}\rm}
 \ifx\answ\bigans\else scaled\magstep1\fi
\ifx\answ\bigans\def\abstractfont{\tenpoint}\else
\font\absit=cmti10 scaled \magstep1
\font\abssl=cmsl10 scaled \magstep1
\font\absrm=cmr10 scaled\magstep1 \font\absrms=cmr7 scaled\magstep1
\font\absrmss=cmr5 scaled\magstep1 \font\absi=cmmi10 scaled\magstep1
\font\absis=cmmi7 scaled\magstep1 \font\absiss=cmmi5 scaled\magstep1
\font\abssy=cmsy10 scaled\magstep1 \font\abssys=cmsy7 scaled\magstep1
\font\abssyss=cmsy5 scaled\magstep1 \font\absbf=cmbx10 scaled\magstep1
\skewchar\absi='177 \skewchar\absis='177 \skewchar\absiss='177
\skewchar\abssy='60 \skewchar\abssys='60 \skewchar\abssyss='60
\def\abstractfont{\def\rm{\fam0\absrm}% switch to abstract font
\textfont0=\absrm \scriptfont0=\absrms \scriptscriptfont0=\absrmss
\textfont1=\absi \scriptfont1=\absis \scriptscriptfont1=\absiss
\textfont2=\abssy \scriptfont2=\abssys \scriptscriptfont2=\abssyss
\textfont\itfam=\absit \def\it{\fam\itfam\absit}\def\footnotefont{\tenpoint}%
\textfont\slfam=\abssl \def\sl{\fam\slfam\abssl}%
\textfont\bffam=\absbf \def\bf{\fam\bffam\absbf}\rm}\fi
\def\tenpoint{\def\rm{\fam0\tenrm}% switch back to 10-point type
\textfont0=\tenrm \scriptfont0=\sevenrm \scriptscriptfont0=\fiverm
\textfont1=\teni  \scriptfont1=\seveni  \scriptscriptfont1=\fivei
\textfont2=\tensy \scriptfont2=\sevensy \scriptscriptfont2=\fivesy
\textfont\itfam=\tenit \def\it{\fam\itfam\tenit}\def\footnotefont{\ninepoint}%
\textfont\bffam=\tenbf \def\bf{\fam\bffam\tenbf}\def\sl{\fam\slfam\tensl}\rm}
\font\ninerm=cmr9 \font\sixrm=cmr6 \font\ninei=cmmi9 \font\sixi=cmmi6
\font\ninesy=cmsy9 \font\sixsy=cmsy6 \font\ninebf=cmbx9
\font\nineit=cmti9 \font\ninesl=cmsl9 \skewchar\ninei='177
\skewchar\sixi='177 \skewchar\ninesy='60 \skewchar\sixsy='60
\def\ninepoint{\def\rm{\fam0\ninerm}% switch to footnote font
\textfont0=\ninerm \scriptfont0=\sixrm \scriptscriptfont0=\fiverm
\textfont1=\ninei \scriptfont1=\sixi \scriptscriptfont1=\fivei
\textfont2=\ninesy \scriptfont2=\sixsy \scriptscriptfont2=\fivesy
\textfont\itfam=\ninei \def\it{\fam\itfam\nineit}\def\sl{\fam\slfam\ninesl}%
\textfont\bffam=\ninebf \def\bf{\fam\bffam\ninebf}\rm}
%
%---------------------------------------------------------------------
%
\def\noblackbox{\overfullrule=0pt}
\hyphenation{anom-aly anom-alies coun-ter-term coun-ter-terms}
\def\inv{^{\raise.15ex\hbox{${\scriptscriptstyle -}$}\kern-.05em 1}}

\def\Dsl{\,\raise.15ex\hbox{/}\mkern-13.5mu D} %this one can be subscripted
\def\dsl{\raise.15ex\hbox{/}\kern-.57em\partial}

 %pound sterling
\def\lspace{\ifx\answ\bigans{}\else\qquad\fi}
\def\lbspace{\ifx\answ\bigans{}\else\hskip-.2in\fi} % $$\lbspace...$$
\def\boxeqn#1{\vcenter{\vbox{\hrule\hbox{\vrule\kern3pt\vbox{\kern3pt
	\hbox{${\displaystyle #1}$}\kern3pt}\kern3pt\vrule}\hrule}}}
\def\mbox#1#2{\vcenter{\hrule \hbox{\vrule height#2in
		\kern#1in \vrule} \hrule}}  %e.g. \mbox{.1}{.1}
%	matters of taste
%\def\tilde{\widetilde} \def\bar{\overline} \def\hat{\widehat}
%
% some sample definitions
  %     curly letters

\def\vev#1{\langle #1 \rangle}

\def\darr#1{\raise1.5ex\hbox{$\leftrightarrow$}\mkern-16.5mu #1}
 %pound sterling

 %puts a small half in a displayed eqn
\def\roughly#1{\raise.3ex\hbox{$#1$\kern-.75em\lower1ex\hbox{$\sim$}}}

%%%%%%%%%%%%%%%%%%%%%%%%%%%%%%%%%%%%%%%%%%%%%%%%%%%%%%%%%%%%%%%%%%%%%
%%%%%%%%%%%%%%%   Subsubsection  %%%%%%%%%%%%%%%%%%%%%%%%%%%%%%%%%%%%
%%%%%%%%%%%%%%%%%%%%%%%%%%%%%%%%%%%%%%%%%%%%%%%%%%%%%%%%%%%%%%%%%%%%%
\global\newcount\subsubsecno \global\subsubsecno=0
\def\subsubsec#1{\global\advance\subsubsecno by1%
{\toks0{#1}\message{(\the\secno\the\subsecno\the\subsubsecno. \the\toks0)}}%
\ifnum\lastpenalty>9000\else\bigbreak\fi
\noindent{\it\hyperdef\hypernoname{subsubsection}{\the\secno.\the\subsecno\the\subsubsecno}%
{\the\secno.\the\subsecno.\the\subsubsecno.} #1}
%%% Add Subsubsections to Index
%% \writetoca{\string\quad{\string\hyperref{}{subsubsection}{\the\secno\the\subsecno\the
%%\subsubsecno}{\baselineskip=9pt\it\the\secno.\the\subsecno.\the\subsubsecno.}}
%% {\baselineskip=9pt\it\ #1}}
\par\nobreak\medskip\nobreak}
%%%%%%%%%%%%%%%%%%%%%%%%%%%%%%%%%%%%%%%%%%%%%%%%%%%%%%%%%%%%%%%%%%%%%
%%%%%%%%%%%%%%%%%%%%%%%%%%%%%%%%%%%%%%%%%%%%%%%%%%%%%%%%%%%%%%%%%%%
%%%%%% BOX
%%%%%%%%%%%%%%%%%%%%%%%%%%%%%%%%%%%%%%%%%%
\def\boxit#1{\vbox{\hrule\hbox{\vrule\kern8pt
\vbox{\hbox{\kern8pt}\hbox{\vbox{#1}}\hbox{\kern8pt}}
\kern8pt\vrule}\hrule}}
\def\mathboxit#1{\vbox{\hrule\hbox{\vrule\kern8pt\vbox{\kern8pt
\hbox{$\displaystyle #1$}\kern8pt}\kern8pt\vrule}\hrule}}
%%%%%%%%%%%%%%%%%%%%%%%%%%%%%%%%%%%%%%%%%%%%%%%%%%%%%%%%%%%%%%%%%%%
%%%%%%%%%%%%%%%%%%%%%%%%%%%%%%%%%%%%%%%%%%%%%%%%%%%%%%%%%%%%%%%%
%%%%%   Dirac-Slash
%%%%%%%%%%%%%%%%%%%%%%%%%%%%%%%%%%%%%%%%%%%%%%%%%%%%%%%%%%%%%%%%
\def\slashchar#1{\setbox0=\hbox{$#1$}           % set a box for #1
   \dimen0=\wd0                                 % and get its size
   \setbox1=\hbox{/} \dimen1=\wd1               % get size of /
   \ifdim\dimen0>\dimen1                        % #1 is bigger
      \rlap{\hbox to \dimen0{\hfil/\hfil}}      % so center / in box
      #1                                        % and print #1
   \else                                        % / is bigger
      \rlap{\hbox to \dimen1{\hfil$#1$\hfil}}   % so center #1
      /                                         % and print /
   \fi}
%%%%%%%%%%%%%%%%%%%%%%%%%%%%%%%%%%%%%%%%%%%%%%%%%%%%%%%%%%%%%%%%%
%%%%%%%%%%%%%%%%%%%%%%%%%%%%%%%%%%%%%%%%%%%%%%%%%%%%%%%%%%%
%  To produce a box for a Dalembertian (adapted from p. 320 of TeXbook):
\def\sqr#1#2{{\vcenter{\vbox{\hrule height.#2pt
         \hbox{\vrule width.#2pt height#1pt \kern#1pt
            \vrule width.#2pt}
         \hrule height.#2pt}}}}

%%%%%%%%%%%%%%%%%%%%%%%%%%%%%%%%%%%%%%%%%%%%%%%%%%%%%%%%%%%